%% file: rock1907.tex
\documentclass[conference]{IEEEtran}
\IEEEoverridecommandlockouts
\usepackage{cite}
\usepackage{amsmath,amssymb,amsfonts}
\usepackage{algorithmic}
\usepackage{graphicx}
\usepackage{textcomp}
\usepackage{xcolor}
\usepackage{subfigure}
\usepackage{todonotes}
\usepackage{mathtools}
\usepackage{array}
\usepackage{caption}
\usepackage{notoccite}
\usepackage{pdfpages}

\usepackage[utf8]{inputenc}
\usepackage{pgfplots}
\usetikzlibrary{calc}
\usetikzlibrary{fit,backgrounds}

\pgfdeclarelayer{background}
\pgfdeclarelayer{front}
\pgfsetlayers{background,main,front}

\makeatletter
\pgfkeys{%
	/tikz/on layer/.code={
		\pgfonlayer{#1}\begingroup
		\aftergroup\endpgfonlayer
		\aftergroup\endgroup
	},
	/tikz/node on layer/.code={
		\gdef\node@@on@layer{%
			\setbox\tikz@tempbox=\hbox\bgroup\pgfonlayer{#1}\unhbox\tikz@tempbox\endpgfonlayer\egroup}
		\aftergroup\node@on@layer
	},
	/tikz/end node on layer/.code={
		\endpgfonlayer\endgroup\endgroup
	}
}

\def\node@on@layer{\aftergroup\node@@on@layer}

\pgfplotsset{ every non boxed x axis/.append style={x axis line style=-},
	every non boxed y axis/.append style={y axis line style=-}}

\def\BibTeX{{\rm B\kern-.05em{\sc i\kern-.025em b}\kern-.08em
    T\kern-.1667em\lower.7ex\hbox{E}\kern-.125emX}}

\begin{document}

\title{Complex Signal Denoising and Interference Mitigation for Automotive Radar Using Convolutional Neural Networks\\
\thanks{This work was supported by the Austrian Research Promotion Agency (FFG) under the project SAHaRA (17774193) and NVIDIA by providing GPUs.}
}

\author{\IEEEauthorblockN{Johanna Rock$^{1}$, Mate Toth$^{1,2}$, Elmar Messner$^1$, Paul Meissner$^2$,  Franz Pernkopf$^1$}
	\IEEEauthorblockA{$^1$Graz University of Technology, Austria}
	\IEEEauthorblockA{$^2$Infineon Technologies Austria AG, Graz}
	Email: johanna.rock@tugraz.at}
\maketitle

\begin{abstract}
	\input{abstract.tex}
\end{abstract}


\input{introduction.tex}

\input{signalmodel.tex}


\section{Methodology}
\label{sec:methodology}

The proposed denoising and interference mitigation network architecture is based on CNNs. We investigate two different denoising approaches, RPD and RDD, as shown in Fig.~\ref{fig:spchain_block}. Furthermore, we use either one input channel for the log-magnitude spectrogram (LMS) or two channels, i.e. the real and imaginary parts of the complex-valued spectrogram (RIS). The underlying goal is the same, namely, to reduce the impact of both noise and interference in order to enable a reliable detection of object parameters at a large sensitivity.

\subsection{Model architecture}

{
\thinmuskip=0mu
\medmuskip=0mu
\tiny

\begin{figure*}
	\centering
	\includegraphics[width=\textwidth]{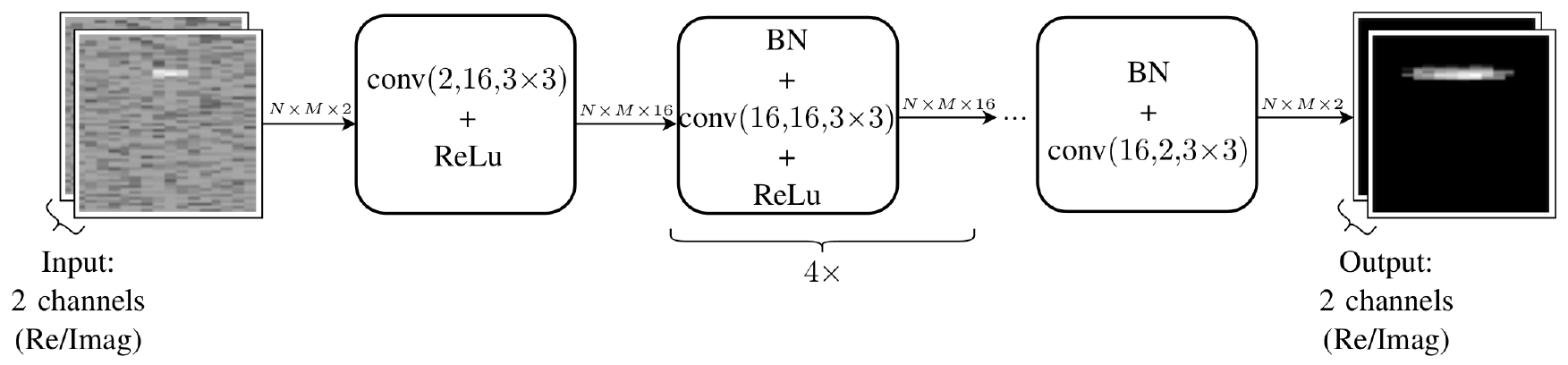}
	\caption{Proposed CNN architecture for radar signal denoising. It uses \emph{ReLu}, \emph{Batch Normalization (BN)} and the convolution operation $\textrm{conv}(i,o,s_1\times s_2)$; see Fig.~\ref{fig:cnn_arch_conv} for details on the convolution operation.}
	\label{fig:cnn_arch}
\end{figure*}
}

\begin{figure}
	\centering
	\includegraphics[width=0.9\columnwidth]{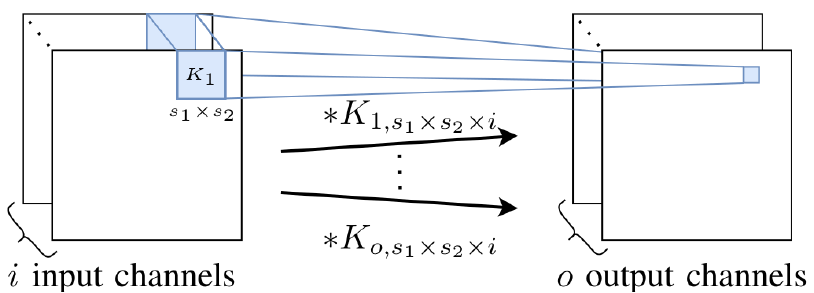}
	\caption{Schematic representation of the convolution operation $\textrm{conv}(i,o,s_1\times s_2)$, where $i$ is the number of input channels, $o$ is the number of kernels, and thus output channels, and $s_1\times s_2$ is the kernel size. The convolution with a kernel is indicated by $*$.}
	\label{fig:cnn_arch_conv}
\end{figure}

The proposed CNN architecture (see Figs.~\ref{fig:cnn_arch} and ~\ref{fig:cnn_arch_conv}) consists entirely of convolutional layers. The first layer uses the convolution operation and a \emph{ReLu} \cite{journals/jmlr/GlorotBB11} activation function, while subsequent layers include \emph{Batch Normalization} \cite{DBLP:journals/corr/IoffeS15}, the convolution operation and the \emph{ReLu} non-linearity, except the last layer which uses a linear activation function instead.

The model architecture differs slightly for the two approaches RPD and RDD. RPD is applied to one-dimensional data ($1 \times N$) and therefore uses one-dimensional kernels. RDD has two-dimensional input samples, i.e. $N \times M$ patches, and uses square kernels. In both approaches, zero-padding is used for the values at the outer boundaries, such that the inputs and outputs for each layer have the same dimensionality.

\subsection{LMS versus RIS Denoising}

We compare denoising with LMS and RIS inputs with focus on performance, memory requirements and application relevance. Denoising LMS can be used for object detection tasks but not for further processing based on the complex spectral values, because the phase information is lost. In RD LMS denoising, the original data, which was not processed by the NN, can be used for further AS calculations. With RIS inputs on the other hand, the denoised spectra can also be directly used for further processing, thus for the RD calculation in RP denoising or the AS calculations in RD denoising.

RDD with e.g. six layers and sixteen kernels with size $3 \times 3$  requires 10002 parameters for RIS and 9713 parameters for LMS inputs, this corresponds to a reduction of only 2.8 \%. A comparable architecture for RPD, i.e. eight layers and eight kernels with size $1 \times 41$, requires 17210 parameters for RIS denoising. See Section \ref{sec:experimental-results} for a more detailed performance comparison.

\subsection{Data Preprocessing}

Prior to model training and evaluation the data samples are standardized, to increase learning capability and model robustness. Two methods are investigated: \emph{Zero-Mean Unit-Variance Scaling (ZMUVS)} and \emph{Complex Standard Scaling (CSS)} \cite{DBLP:journals/corr/TrabelsiBSSSMRB17}.
In both approaches the complex data points are translated to zero-mean, while for ZMUVS the data points are then scaled to unit-variance and for CSS the data points are scaled to the standard normal complex distribution using the inverse square root of the covariance matrix of real and imaginary values.

\subsection{Loss functions}
\label{sec:learn-params}
The loss function defines the similarity of NN-outputs to the NN-targets, thus represents the learning goal. We evaluate the following measures:

\begin{itemize}
	\item The \emph{Mean Squared Error (MSE)} is calculated from the real and imaginary parts of the values of the complex spectrogram.
	\item The \emph{Signal-to-interference-plus-noise-ratio (SINR)} is the proportion of signal power compared to the noise floor, where the latter is given by both, noise and interference.
	\item The \emph{Weighted MSE} is determined as convex combination of the MSE of the complex spectrum, and the magnitude and phase of object peaks.
\end{itemize}

\subsection{Training Setup}
The \emph{Adam} \cite{DBLP:journals/corr/KingmaB14} algorithm is used for training with a learning rate of $0.00005$ and two input samples per batch.


\section{Experimental Setup}
\label{sec:experimental-setup}

In our experiments, we use simulated FMCW/CS radar signals, which gives us access to interfered data and their corresponding clean equivalent. The basic receive IF signal is generated according to~\eqref{eq:signal-model} and processed as described in Section~\ref{sec:sigmod}. The resulting signals depend on the parameters of the random scenarios, which are generated according to uniform distributions $\mathcal{U}(min,max)$ in the respective domains. Among these parameters are the number of objects $\mathcal{U}(1,20)$ and for each object the distance $\mathcal{U}(0\mathrm{m},153\mathrm{m})$ and velocity $\mathcal{U}(-20\mathrm{m/s},20\mathrm{m/s})$ relative to the radar, such that all object parameters lie within the radar's limits.

The interferer parameters are uniformly sampled within the ranges depicted in Table \ref{tab:interferer-param}. The SIR and SNR are used to scale the interference and noise powers relative to the object signal power respectively, when generating the interfered and noisy time domain signal $s_{IF}[n,m]$. The victim radar parameters are kept constant and chosen as shown in Table~\ref{tab:victim-radar-param}. Fig.~\ref{fig:rd-interfered-and-target} shows a RD map processed from simulated data from a scenario with eight objects, where Fig.~\ref{fig:rd-interfered} shows an interfered signal and Fig.~\ref{fig:rd-target} shows the corresponding clean data.

\begin{table}
	\centering
	\vspace{0.5cm}
	\caption{Ranges of interference and noise parameters.}
	\begin{tabular}{l l r r}
		Parameter & & Lower limit & Upper limit \\
		\hline
		$N_\textrm{I}$ & Number of interferers & 1 & 3 \\
		$f_{\textrm{0,I}}$ & Sweep start frequency & $75.8 \textrm{GHz}$ & $76.2 \textrm{GHz}$ \\
		$B_\textrm{I}$ & Sweep bandwidth & $0.6 \textrm{GHz}$ & $1.4 \textrm{GHz}$ \\
		$T_\textrm{I}$ & Sweep duration & $40 \mu \textrm{s}$ & $46 \mu \textrm{s}$ \\
		$\textrm{SIR}$ & Signal-to-interference-ratio & $-20 \textrm{dB}$ & $-60 \textrm{dB}$ \\
		$\textrm{SNR}$ & Signal-to-noise-ratio & $-10 \textrm{dB}$ & $+10 \textrm{dB}$ \\
	\end{tabular}
	\label{tab:interferer-param}
\end{table}

\begin{table}
	\centering
	\vspace{0.5cm}
	\caption{Victim radar and signal processing parameters.}
	\begin{tabular}{l l r r}
		Parameter & & Value \\
		\hline
		$f_{\textrm{0,V}} $ & Sweep start frequency & $76 \textrm{GHz}$ \\
		$B_\textrm{V}$ & Sweep bandwidth & $1 \textrm{GHz}$ \\
		$T_\textrm{V}$ & Sweep duration & $48 \mu \textrm{s}$ \\
		$B_{\textrm{IF,V}}$ & IF bandwidth & $20 \textrm{MHz}$ \\
		$N$ & Number of fast-time samples & $1024$ \\
		$M$ & Number of slow-time samples/ ramps & $128$ \\
		$A$ & Number of antennas & $8$ \\
		$w$ & Window type & $\textrm{Hann}$ \\
	\end{tabular}
	\label{tab:victim-radar-param}
\end{table}

\begin{figure}
	\footnotesize
	\centering
	\subfigure[Interfered]{
		\input{eval_evaluation_test_rd_doppler-range_matrix_interfered_p8_id0.tex}
		\label{fig:rd-interfered}
	}
	\hspace{-0.6cm}
	\subfigure[Clean]{
		\input{eval_evaluation_test_rd_doppler-range_matrix_targets_p8_id0.tex}
		\label{fig:rd-target}
	}
	\caption{Exemplary range-Doppler magnitude spectra in dB of a scenario with eight objects.}
	\label{fig:rd-interfered-and-target}
\end{figure}

\subsection{Data Sets}

Three separate data sets are used for training, validation and testing the models. The data sets contain samples of 2000 scenarios, i.e. RPs or RD maps, for training, and 250 scenarios for validation and testing each. Data from a single scenario are exclusively contained either in the training, validation or test set.

\subsection{Performance Measures}
\label{subsec:perfmeas}

For performance evaluation, we examine different quantitative and qualitative measures, which cover two fundamental aspects of object detection in chirp sequence radar processing \cite{performance-comparison-interf-mitigation}:

\begin{itemize}
	\item The \emph{detection probability} gives the chance that an object is detected on the RD map.
	\item The \emph{determination of detected object properties} is defined through the correctness of object location on the RD map as well as object resolution and peak distortion, which provide information about the object's radar cross section and thus its physical characteristics.
\end{itemize}

The goal of interference mitigation is to maximize the detection probability while avoiding modifications in object properties, i.e. the object peak's magnitude and phase.

\subsubsection{Quantitative measures}

The \emph{signal-to-interference-plus-noise ratio (SINR)} directly relates to the detection probability. It is defined through the ratio of signal power at the object peaks compared to the noise floor. In the two-dimensional case, i.e. in the range-Doppler domain, for multi-object scenarios the SINR is defined as:
\begin{equation}
\textrm{SINR} = 10 \log \bigg(\frac{
	\frac{1}{N_O} \sum_{\{n,m\} \in \mathcal{O}}^{} {\mid \tilde{S}_\textrm{RD}[n,m]\mid}^{2}
} {
	\frac{1}{N_N} \sum_{\{n,m\} \in \mathcal{N}}^{} {\mid \tilde{S}_\textrm{RD}[n,m]\mid}^{2}
}\bigg),
\label{sinr}
\end{equation}
where $n$ and $m$ are row and column indices of the RD matrix, $\mathcal{O}$ is the set of object peaks and $\mathcal{N}$ is the set of $N_N$ noise cells. Noise cells are defined to have a minimum distance to each object peak depending on the bin width in distance and velocity domain as well as the physical resolution of the radar. In the one-dimensional case, thus for AS evaluation, the SINR is defined analogously.

The \emph{error vector magnitude (EVM)} gives information about the detected object properties. It is defined as the magnitude of the error vector between the clean RD map $S_\textrm{RD,clean}$ and the denoised signal $\tilde{S}_\textrm{RD}$, i.e. in a multi-object scenario:

\begin{equation}
\textrm{EVM} = \frac{1}{N_O} \sum_{\{n,m\} \in \mathcal{O}}^{}\frac{\mid S_\textrm{RD,clean}[n,m] - \tilde{S}_\textrm{RD}[n,m]\mid}{\mid S_\textrm{RD,clean}[n,m] \mid}.
\end{equation}

\subsubsection{Qualitative measures}

During visual inspection of the RD map and the AS, we consider criteria such as object peak and noise floor intensity, object peak location, resolution and distortion as well as artifact appearances.

\input{setup_comparison.tex}


\section{Experimental Results}
\label{sec:experimental-results}
The proposed NN-based architecture is evaluated in several steps: First, the optimal network architectures for RIS denoising are analyzed using the MSE loss function and ZMUVS as ``basic" setup. The best performing architectures in terms of overall-performance and performance-complexity are further evaluated with respect to scaling methods, loss functions and according to their generalization capabilities. Second, the different proposed CNN-based approaches, i.e. RPD and RDD with LMS and RIS inputs, are evaluated and analyzed. Third, the best-performing CNN-based model is compared to state-of-the-art mitigation algorithms.

\subsection{Analysis of Optimal CNN-Architecture}
\label{sec:exp-results-architecture}

We used grid search to systematically find the best CNN architecture for RD RIS denoising, i.e. the number of layers ($4$, $6$, $8$), kernel sizes ($1\times1$, $3\times3$, $5\times5$, $7\times 7$) and number of kernels per layer ($2$, $8$, $16$, $32$). The choice of the kernel size is a trade-off between object resolution in the denoised spectrogram and noise suppression. Larger kernels enable better denoising performance, but result in distortion of the peak shapes and thus a possible loss of resolution. We use ZMUVS and the MSE loss function for training the CNNs.

Fig.~\ref{fig:arch-comparison} shows the SINR and EVM based performance comparison of all evaluated RD architectures for RIS inputs using MSE loss and ZMUVS. The SINR and EVM are illustrated in blue and red respectively, while the x-axis indicates the number of parameters of the NN-model. The best performing models are marked with A to F and listed in detail in Table \ref{tab:arch-comparison-top-results}. The best performance for RD RIS denoising (Model D) is obtained using a model with $6$ layers, a kernel size of $3\times 3$ and $16$ kernels per layer. Model A offers the best performance-complexity trade-off with only four layers and two kernels with a size of $3\times 3$. With an average SINR loss of only 3.8 dB, there is a parameter reduction of 98.4 \% compared to Model D. However, the small model size comes with the cost of a high average EVM. This indicates a large distortion of the complex values of object peaks, which causes a notable decrease of SINR in the AS. For the other architectures listed in Table~\ref{tab:arch-comparison-top-results} the EVM values appear to be small enough such that other effects, e.g. the main- to side-lobe ratio of the AS peak, dominate the AS SINR.

Using similar parameter ranges except for one-dimensional kernels ($5$, $13$, $21$, $25$, $31$, $41$, $43$, $47$, $51$, $55$, $61$), the most suitable architecture for RP denoising has $6$ layers, a kernel size of $1\times 41$ and $16$ kernels per layer.

\begin{figure}
	\centering
	\resizebox {\columnwidth} {!} {
		\input{arch_comparison.tex}
	}
	\caption{CNN architecture performance comparison for RD RIS denoising using MSE loss and ZMUVS. Labels A-F refer to model parameters given in Table~\ref{tab:arch-comparison-top-results}.}
	\label{fig:arch-comparison}
\end{figure}

\begin{table*}
	\centering
	\vspace{0.5cm}
	\caption{Best performing architectures for RD RIS denoising.}
	\begin{tabular}{l r r r r >{\bfseries}r r >{\bfseries}r r}
		Model & Layers & Kernels & Kernel Size & Parameters & SINR (RD) & EVM (RD) & SINR (AS) \\
		\hline
		A & 4 & 2 & ($3 \times 3$) & 160 & 73.67 & 0.90 & 7.50 \\
		B & 8 & 8 & ($3 \times 3$) & 3898 & 73.60 & 0.40 & 9.47 \\
		C & 4 & 16 & ($3 \times 3$) & 5298 & 74.57 & 0.30 & 9.39 \\
		D & 6 & 16 & ($3 \times 3$) & 10002 & 77.47 & 0.58 & 9.56 \\
		E & 8 & 16 & ($3 \times 3$) & 14706 & 77.78 & 0.47 & 9.54 \\
		F & 6 & 32 & ($3 \times 3$) & 38434 & 72.20 & 0.51 & 9.43 \\
	\end{tabular}
	\label{tab:arch-comparison-top-results}
\end{table*}

Data scaling has a strong impact on the training progress in terms of duration and stability. While CSS statistically results in stronger denoising performance when using MSE loss (on average additional 11.41 dB SINR), it does not seem to have any positive effects when using SINR loss. CSS has a slightly negative effect on the object peak values which results in an increased average EVM. ZMUVS on the other hand produces results with a lower average SINR, but it also has a lower average EVM. Additionally, CSS leads to a smoother and more stable learning improvement.

When analyzing the loss functions introduced in Section \ref{sec:learn-params}, we can see that the SINR loss produces the highest denoising performance for most models, but it fails to preserve phase information for further processing. For example when evaluating Model D with ZMUVS, it increases the SINR performance metric by an average of 17.58 dB compared to the MSE loss function. On the contrary the MSE loss takes the relation of imaginary and real values of the inputs into account, and thus better preserves object peak values. The \emph{Weighted MSE} performs worse than the other loss functions, both in terms of SINR and EVM.

We use a second pair of training and validation data sets, that contain only up to two interferer, in order to investigate generalization capabilities. Thus we want to show how the trained network performs on a test set coming from a slightly different distribution, i.e. with three interferers, than the data seen during the training process. With an average SINR loss of 5.6 dB (Model A) and 23.89 dB (Model D) RD RIS denoising seems to generalize to similar data as seen during training. Even with a decreased SINR by 23.89 dB RD RIS denoising outperforms the state of the art. However, the performance loss between the two models suggests that stronger regularization is required for bigger architectures.

\subsection{Performance Analysis of CNN-based Approaches}
\label{sec:performance-cnn}

The different CNN-based approaches as described in Section~\ref{sec:methodology} are analyzed using performance metrics as introduced in Section~\ref{subsec:perfmeas}. For RD LMS denoising and RD RIS denoising, we use Models A and D from Section~\ref{sec:exp-results-architecture}, thus the best architectures in performance-complexity and overall-performance respectively. RPD is performed using the model described in Section~\ref{sec:exp-results-architecture}. We use CCS and the MSE loss function for training the CNNs.

The performance is illustrated using the cumulative distribution function (CDF) of the respective metric. The values are computed from the test set in a Monte Carlo simulation as introduced in Section~\ref{sec:experimental-setup}. The interfered signal without mitigation (\emph{interfered}) and the signal with only AWGN (\emph{noisy}) are included as references.

\begin{figure}[tb]
	\centering
	\resizebox {0.9\columnwidth} {!} {
		\input{rd-sinr_cdf_cnns.tex}
	}
	\caption{CDF of RD SINR of different CNN-based models.}
	\label{fig:cdf-rd-sinr-cnn}
\end{figure}

Fig.~\ref{fig:cdf-rd-sinr-cnn} shows the range-Doppler SINR performance of the different models. RD LMS denoising generally requires larger models in order to perform well. Model A with LMS fails to learn the denoising task, while Model D with LMS seems to perform quite well on average. However, the SINR CDF and visual inspections suggest that RD LMS denoising performs well on data with weak interference while the denoising performance drastically decreases on data with stronger interference. In severe cases, it completely fails to detect object peaks contained in the RD map. Thus, this approach is rather unreliable and performs worse compared to the other CNN-based approaches.

RPD has a lower average SINR than the other approaches, but it shows remarkably low variance. This suggests, that RPD is a more stable approach, with a solid performance also on strongly interfered data.

RD RIS denoising results in a superior average SINR compared to the other approaches. Model A performs well on around 75 percent of the data samples despite its small model size. The SINR drops notably for the other 25 percent tough, which suggests that Model A is not capable of denoising a broad variety of interference. Model D on the other hand has an even stronger and also more reliable denoising performance than Model A, which becomes apparent especially for scenarios with stronger interference.

\input{results_comparison.tex}


\input{conclusion.tex}

\bibliographystyle{ieeetr}
\bibliography{bibliography}

\end{document}

%% file: abstract.tex
Driver assistance systems as well as autonomous cars have to rely on sensors to perceive their environment. A heterogeneous set of sensors is used to perform this task robustly. Among them, radar sensors are indispensable because of their range resolution and the possibility to directly measure velocity. Since more and more radar sensors are deployed on the streets, mutual interference must be dealt with. In the so far unregulated automotive radar frequency band, a sensor must be capable of detecting, or even mitigating the harmful effects of interference, which include a decreased detection sensitivity.
In this paper, we address this issue with Convolutional Neural Networks (CNNs), which are state-of-the-art machine learning tools. We show that the ability of CNNs to find structured information in data while preserving local information enables superior denoising performance. To achieve this, CNN parameters are found using training with simulated data and integrated into the automotive radar signal processing chain. The presented method is compared with the state of the art, highlighting its promising performance. Hence, CNNs can be employed for interference mitigation as an alternative to conventional signal processing methods.
Code and pre-trained models are available at \emph{https://github.com/johanna-rock/imRICnn}.

%% file: introduction.tex
\section{Introduction}

Automotive radar sensors are key elements of current driving assistance systems as well as of autonomous driving applications. Nowadays, frequency modulated continuous wave (FMCW)/chirp sequence (CS) radars are prevalent. They share a non-regulated spectrum, transmitting sequences of linear chirp signals. Requirements for fine range resolution demand ever larger radio frequency (RF) transmit bandwidths, while the number of sensors deployed is also rising. Hence, mutual interference between radar sensors is becoming increasingly likely. The most common form of mutual interference is non-coherent interference~\cite{TOT18}, in which radar sensors with non-identical transmit signal parameters interfere. This leads to time-limited broadband disturbances in the baseband signal, whose primary effect is a reduced object detection sensitivity~\cite{BRO07}. Therefore, interference mitigation is a crucial part of current and future radar sensors used in a safety context.

Several conventional signal processing algorithms have been proposed in order to mitigate mutual interference. The most basic method is to zero out detected interference samples. In \cite{WAG18}, nonlinear filtering in slow-time is performed to remove interference. A different method is proposed in \cite{Bechter2017a}, where the useful signal is iteratively reconstructed using Fourier transforms and thresholding \cite{MAR12}. The interference component of the signal may be estimated and subtracted, such as in~\cite{BEC17}. Furthermore, beamforming can be used to reduce the impact of interference from particular directions \cite{Bechter2016}. Some machine learning techniques were discussed in the context of interference detection and classification in~\cite{Zhang2018a}. However, no explicit machine learning approach has been proposed in the context of interference mitigation.

In this paper, we use neural networks (NNs) as a powerful machine learning method to mitigate interference. In particular, convolutional NNs (CNNs) are employed. They are capable of learning local patterns, by considering inputs that are located close-by, and recognize them throughout the whole data signal. This structure can also be advantageous for spectrogram representations. Additionally, CNNs require a relatively small amount of learnable parameters compared to fully connected NNs, which makes them more appropriate for deployment on resource-constrained systems such as integration on chip level.
We will show how a two-channel representation of complex spectrogram data~\cite{fu2017complex} can be used as network input at two different points in the processing chain.

Since automotive radar is a safety-critical application, certain requirements must be fulfilled for interference mitigation and signal denoising algorithms. Besides an adequate noise suppression, no artifacts may be generated by the processing that can lead to spurious detections (ghost objects). We address these issues by using a detailed performance comparison that evaluates different application-relevant measures in a Monte-Carlo simulation~\cite{performance-comparison-interf-mitigation}.

Main contributions of this paper are:
\begin{itemize}
  \item We show specific CNN structures capable of denoising radar signals.
  \item We present numerical results using application-related performance metrics in a comparison with the state of the art.
  \item We show that an excellent level of noise reduction and hence an improvement of detection sensitivity can be achieved.
\end{itemize}

%% file: signalmodel.tex
\section{Signal Model}
\label{sec:sigmod}

\begin{figure}[tb]
	\centering
	\input{spchain.tex}
	\caption{Block diagram of a basic FMCW/CS radar processing chain. Dashed boxes indicate the locations of optional interference mitigation steps, including our proposed methods. The signal at every point in the chain is labeled according to the variable names used in this paper.}
	\label{fig:spchain_block}
\end{figure}

Fig.~\ref{fig:spchain_block} illustrates the \emph{range-Doppler (RD) processing} chain in a conventional FMCW/CS radar. First, the radar sensor performs a measurement by transmitting a sequence of linearly RF modulated chirp signals, also called \emph{ramps}. For each ramp, the received object reflections are then mixed, i.e. multiplied, with the transmit signal. This way the according time delays, and hence range information, are translated into corresponding constant frequency sinusoidals. The signal after mixing is limited by the \emph{intermediate frequency (IF)} bandwidth of the receiver and therefore termed the \emph{IF signal}. The velocity (Doppler) information is estimated by evaluating the rate of ramp-wise linear phase change of the received IF signal. More detailed descriptions as well as mathematical derivations of these principles can be found in~\cite{STO92,WIN07}.

From a data processing point of view, the received IF signal consists of $N$ time domain samples for each of the $M$ transmitted ramps. Hence, it can be interpreted as a two-dimensional data matrix $s_{\mathrm{IF}}[n,m]$ with the corresponding indices $n$ and $m$, also called \emph{fast-} and \emph{slow-time}, respectively. Essential processing steps include discrete Fourier transforms (DFTs) over the fast and slow-time, to reveal distance and velocity information accordingly. The resulting two-dimensional spectrum $S_{\mathrm{RD}}[n,m]$ ideally contains peaks at the objects' corresponding distances and velocities, which are then to be detected. In a system with multiple receive channels (antennas), additional information about the angle of arrival of object reflections can be extracted. This is done by evaluating the phase change of each object peak value over the receive channels by another DFT, yielding the so-called \emph{angular spectrum (AS)}. Further processing steps are then performed in higher layers of the application and may include sensor fusion, tracking,  or classification.

However, in addition to the object reflections, IF signals in real radar systems also contain disturbances in the form of receiver noise and (mutual) interference. Other radar sensors in the radio range act as \emph{interferers} when transmitting inside the receiver IF bandwidth of the \emph{victim radar}. Accounting for this, the model of the IF signal $s_{\mathrm{IF}}[n,m]$ can be written as

\begin{equation}
s_{\mathrm{IF}}[n,m]=\sum_{o=1}^{N_{\mathrm{O}}} s_{\mathrm{O},o}[n,m] + \sum_{i=1}^{N_{\mathrm{I}}} s_{\mathrm{I},i}[n,m] + \upsilon[n,m] \, ,
\label{eq:signal-model}
\end{equation}
where $s_{\mathrm{O},o}[n,m]$ is the signal component of the $o^{th}$ object reflection, $N_{\mathrm{O}}$ denotes the number of objects, $s_{\mathrm{I},i}[n,m]$ is the signal component of the $i^{th}$ interferer assuming $N_{\mathrm{I}}$ interferers, and $\upsilon[n,m]$ is a receiver noise term. Receiver noise is modeled as AWGN, while mutual interference generally causes burst-like disturbances in time domain corresponding to broadband disturbances in frequency domain~\cite{TOT18,Kim2018}.

State-of-the-art (``classical") interference mitigation methods are mostly signal processing algorithms that are applied either on the time domain signal $s_{\mathrm{IF}}[n,m]$ or on the frequency domain signal $S_{\mathrm{R}}[n,m]$ after the first DFT. The two NN-based methods presented in this paper are applied at two different steps in the radar signal processing, i.e.,
\begin{enumerate}
	\item Range-Profile Denoising (RPD)\label{rp-denoising}: Denoising of range-profiles after the first DFT.
	\item Range-Doppler Denoising (RDD)\label{rd-denoising}: Denoising of range-Doppler maps after the second DFT.
\end{enumerate}

%% file: spchain.tex
\usetikzlibrary{shapes,arrows,patterns}


\pgfdeclarelayer{background}
\pgfdeclarelayer{foreground}
\pgfsetlayers{background,main,foreground}

\tikzstyle{block}=[draw, fill=black!10, text width=5em, 
    text centered, minimum height=6em]
\tikzstyle{blockan}=[block, fill=red!20,rounded corners]
\tikzstyle{blockmet}=[block, fill=blue!20,rounded corners, draw=black, dashed, very thick]
\def\blockdist{4}

\begin{tikzpicture}[scale=0.7, transform shape]

\node (adc) [blockan] {Radar Sensor};
\path (adc)+(\blockdist,0) node (tdp) [block] {Time Domain Pre-processing};
\path (tdp)+(\blockdist,0) node (rp) [block] {DFT over $n$ \\ for each $m$};
\path (adc)+(0,-0.7*\blockdist) node (rpd) [blockmet] {RPD};
\path (rpd)+(\blockdist,0) node (rd) [block] {DFT over $m$ \\ for each $n$};
\path (rd)+(\blockdist,0) node (rdd) [blockmet] {RDD};
\path (rpd)+(0,-0.7*\blockdist) node (od) [block] {Object Detection};
\path (od)+(1.15*\blockdist,0) node (aoa) [block] {Angle Estimation};
\path (aoa)+(0.85*\blockdist,0) node (etc) [block] {Further Processing};

\path (adc.-90)+(0,-0.05*\blockdist) node (adc1) [draw=none, fill=none] {};
\path (rp.-90)+(0,-0.05*\blockdist) node (rp1) [draw=none, fill=none] {};
\path (tdp.-90)+(0,-0.05*\blockdist) node (tdp1) [draw=none, fill=none] {};
\path (rd.-90)+(0,-0.05*\blockdist) node (rd1) [draw=none, fill=none] {};
\path (rpd.-90)+(0,-0.05*\blockdist) node (rpd1) [draw=none, fill=none] {};
\path (od.-90)+(0,-0.05*\blockdist) node (od1) [draw=none, fill=none] {};
\path (rdd.-90)+(0,-0.05*\blockdist) node (rdd1) [draw=none, fill=none] {};

\draw[very thick, ->] (adc) -- (tdp) node[midway, above] () {$s_{\mathrm{IF}}[n,m]$} node [midway,below] () {${(N{\times}M)}$};
\draw[very thick, ->] (tdp) -- (rp) node[midway, above] () {$\tilde{s}_{\mathrm{IF}}[n,m]$};
\draw[very thick, -stealth] (rp.-90) -- (rp1.-90) -- (adc1.-90) node[near start, above] () {$S_{\mathrm{R}}[n,m]$} -- (rpd.90);
\draw[very thick, ->] (rpd) -- (rd) node[midway, above] () {$\tilde{S}_{\mathrm{R}}[n,m]$};
\draw[very thick, ->] (rd) -- (rdd) node[midway, above] () {$S_{\mathrm{RD}}[n,m]$};
\draw[very thick, -stealth] (rdd.-90) -- (rdd1.-90) -- (rpd1.-90) node[near start, above] () {$\tilde{S}_{\mathrm{RD}}[n,m]$} -- (od.90);
\draw[very thick, ->] (od) -- (aoa) node[midway, above] () {object peaks};
\draw[very thick, ->] (aoa) -- (etc) node[midway, above] () {objects};

\begin{pgfonlayer}{background}
\path (tdp.135)+(0,0.47) node (a) {};
\path (rp.-45) node (t1) {};
\path (rpd.90) node (t2) {};
\path (t1|-t2)+(0,0) node (b) {};
\path[fill=yellow!20,rounded corners, draw=black!50, dashed] (a) rectangle (b) node[anchor = north, rotate = 0, midway, above, xshift = 0, yshift=33pt] () {\scriptsize{\textit{classical interference processing}}};
\end{pgfonlayer}

\end{tikzpicture}


%% file: eval_evaluation_test_rd_doppler-range_matrix_interfered_p8_id0.tex
\begin{tikzpicture}

\begin{axis}[
scale only axis=true,
width=0.3\columnwidth,
height=0.3\columnwidth,
axis background/.style={fill=white!89.80392156862746!black},
axis line style={white},
point meta max=0,
point meta min=-70,
tick align=outside,
tick pos=left,
x grid style={white},
xlabel={Velocity [m/s]},
xmajorgrids,
xmin=-20.5452054340198, xmax=20.2241865991132,
y grid style={white},
ylabel={Distance [m]},
y label style={at={(0.1,0.5)}},
ymajorgrids,
ymin=0, ymax=153.420417589503
]
\addplot graphics [includegraphics cmd=\pgfimage,xmin=-20.5452054340198, xmax=20.2241865991132, ymin=0, ymax=153.420417589503] {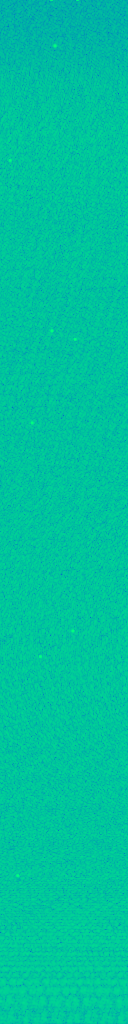};

\end{axis}
\end{tikzpicture}

%% file: eval_evaluation_test_rd_doppler-range_matrix_targets_p8_id0.tex
\begin{tikzpicture}

\begin{axis}[
scale only axis=true,
width=0.3\columnwidth,
height=0.3\columnwidth,
axis background/.style={fill=white!89.80392156862746!black},
axis line style={white},
colorbar,
colorbar style={ylabel={}},
colorbar/width=1.5mm,
colormap={mymap}{[1pt]
  rgb(0pt)=(0,0,1);
  rgb(1pt)=(0,1,0.5)
},
point meta max=0,
point meta min=-70,
tick align=outside,
tick pos=left,
x grid style={white},
xlabel={Velocity [m/s]},
xmajorgrids,
xmin=-20.5452054340198, xmax=20.2241865991132,
y grid style={white},
yticklabels={,,},
ymajorgrids,
ymin=0, ymax=153.420417589503
]
\addplot graphics [includegraphics cmd=\pgfimage,xmin=-20.5452054340198, xmax=20.2241865991132, ymin=0, ymax=153.420417589503] {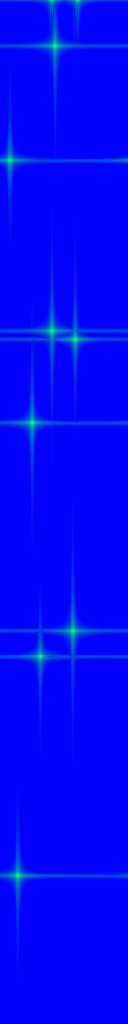};

\end{axis}

\end{tikzpicture}

%% file: setup_comparison.tex
\subsection{Mitigation Methods Selected for Comparison}\label{subsec:setupcomp}

A small number of the most well-known and promising state-of-the-art signal processing algorithms have been chosen for a comparative analysis. This also allows for a discussion of the properties of the NN-based approaches in a broader context of interference mitigation. A short summary of these methods is presented below. 

\subsubsection{Zeroing} Zeroing is selected as a baseline, since it is a simple and well-known method. Time domain samples of the IF signal, that are determined to be dominated by interference, are simply set to a value of zero. Its properties have been discussed in e.g.~\cite{Fischer}.

\subsubsection{Iterative method with adaptive thresholding (IMAT)} IMAT~\cite{Bechter2017a} is based on an initial zeroing step, eliminating interference. The resulting ``gaps" in the signal are then interpolated in an attempt to fully reconstruct the object signal. This reconstruction is done by an iterative thresholding method making use of the theory of sparse sampling. 

\subsubsection{Ramp filtering (RFmin)} Ramp filtering~\cite{WAG18}, as opposed to previously mentioned techniques, processes the signal after the first DFT. It exploits the sparsity and diversity of interference over the slow-time domain, using a non-linear filtering operation to achieve considerable interference as well as noise suppression. Several choices of filtering can be considered. In this work, a simple \emph{minimum operator} is implemented.

Note that both zeroing and IMAT require the detection of interfered IF signal samples. In this paper, it will be assumed that this operation works perfectly. However, in general, errors in interference detection may have a strong impact on the performance of mitigation algorithms~\cite{performance-comparison-interf-mitigation}. Ramp filtering, as well as the proposed novel approaches, are not directly based on an interference detection step.

%% file: arch_comparison.tex
\begin{tikzpicture}

\definecolor{color1}{rgb}{0.886274509803922,0.290196078431373,0.2}
\definecolor{color0}{rgb}{0.203921568627451,0.541176470588235,0.741176470588235}

\begin{axis}[
axis background,
axis line style={white},
tick align=outside,
tick pos=left,
x grid style={line width = 0pt},
xmajorgrids,
xmin=20.2294610720991, xmax=487112.136348054,
xmode=log,
xtick={1,10,100,1000,10000,100000,1000000,10000000},
xticklabels={$10^{0}$,$10^{1}$,$10^{2}$,$10^{3}$,$10^{4}$,$10^{5}$,$10^{6}$,$10^{7}$},
xlabel={Number parameters},
y grid style={line width = 0pt},
ylabel={\textcolor{color0}{SINR [dB]}},
yticklabel style=color0,
ymajorgrids,
ymin=-4.432104862905, ymax=89.606110309205
]
\addplot [semithick, color0, mark=*, mark options={solid}, forget plot]
table [row sep=\\]{%
32	27.5409578822 \\
};
\addplot [semithick, color0, mark=*, mark options={solid}, forget plot]
table [row sep=\\]{%
160	73.6680486383 \\
};
\addplot [semithick, color0, mark=*, mark options={solid}, forget plot]
table [row sep=\\]{%
416	59.9266677766 \\
};
\addplot [semithick, color0, mark=*, mark options={solid}, forget plot]
table [row sep=\\]{%
800	57.6773904145 \\
};
\addplot [semithick, color0, mark=*, mark options={solid}, forget plot]
table [row sep=\\]{%
52	-0.1576405369 \\
};
\addplot [semithick, color0, mark=*, mark options={solid}, forget plot]
table [row sep=\\]{%
244	58.5909080674 \\
};
\addplot [semithick, color0, mark=*, mark options={solid}, forget plot]
table [row sep=\\]{%
628	58.1844080994 \\
};
\addplot [semithick, color0, mark=*, mark options={solid}, forget plot]
table [row sep=\\]{%
1204	57.3168772462 \\
};
\addplot [semithick, color0, mark=*, mark options={solid}, forget plot]
table [row sep=\\]{%
72	-0 \\
};
\addplot [semithick, color0, mark=*, mark options={solid}, forget plot]
table [row sep=\\]{%
328	55.7072683787 \\
};
\addplot [semithick, color0, mark=*, mark options={solid}, forget plot]
table [row sep=\\]{%
840	59.4853093099 \\
};
\addplot [semithick, color0, mark=*, mark options={solid}, forget plot]
table [row sep=\\]{%
1608	59.8816655165 \\
};
\addplot [semithick, color0, mark=*, mark options={solid}, forget plot]
table [row sep=\\]{%
218	66.2288694122 \\
};
\addplot [semithick, color0, mark=*, mark options={solid}, forget plot]
table [row sep=\\]{%
1498	66.4430816339 \\
};
\addplot [semithick, color0, mark=*, mark options={solid}, forget plot]
table [row sep=\\]{%
4058	62.7552543974 \\
};
\addplot [semithick, color0, mark=*, mark options={solid}, forget plot]
table [row sep=\\]{%
7898	62.0441027812 \\
};
\addplot [semithick, color0, mark=*, mark options={solid}, forget plot]
table [row sep=\\]{%
394	56.8334041743 \\
};
\addplot [semithick, color0, mark=*, mark options={solid}, forget plot]
table [row sep=\\]{%
2698	65.1447815661 \\
};
\addplot [semithick, color0, mark=*, mark options={solid}, forget plot]
table [row sep=\\]{%
7306	66.3018892378 \\
};
\addplot [semithick, color0, mark=*, mark options={solid}, forget plot]
table [row sep=\\]{%
14218	62.7459427801 \\
};
\addplot [semithick, color0, mark=*, mark options={solid}, forget plot]
table [row sep=\\]{%
570	54.4928935087 \\
};
\addplot [semithick, color0, mark=*, mark options={solid}, forget plot]
table [row sep=\\]{%
3898	73.5985328336 \\
};
\addplot [semithick, color0, mark=*, mark options={solid}, forget plot]
table [row sep=\\]{%
10554	68.8800019494 \\
};
\addplot [semithick, color0, mark=*, mark options={solid}, forget plot]
table [row sep=\\]{%
20538	65.1940938149 \\
};
\addplot [semithick, color0, mark=*, mark options={solid}, forget plot]
table [row sep=\\]{%
690	61.6360151458 \\
};
\addplot [semithick, color0, mark=*, mark options={solid}, forget plot]
table [row sep=\\]{%
5298	74.5742588344 \\
};
\addplot [semithick, color0, mark=*, mark options={solid}, forget plot]
table [row sep=\\]{%
14514	66.3420960590 \\
};
\addplot [semithick, color0, mark=*, mark options={solid}, forget plot]
table [row sep=\\]{%
28338	63.5896587443 \\
};
\addplot [semithick, color0, mark=*, mark options={solid}, forget plot]
table [row sep=\\]{%
1298	66.8702718026 \\
};
\addplot [semithick, color0, mark=*, mark options={solid}, forget plot]
table [row sep=\\]{%
10002	77.4738937999 \\
};
\addplot [semithick, color0, mark=*, mark options={solid}, forget plot]
table [row sep=\\]{%
27410	63.0885652024 \\
};
\addplot [semithick, color0, mark=*, mark options={solid}, forget plot]
table [row sep=\\]{%
53522	65.6939642661 \\
};
\addplot [semithick, color0, mark=*, mark options={solid}, forget plot]
table [row sep=\\]{%
1906	57.2404149553 \\
};
\addplot [semithick, color0, mark=*, mark options={solid}, forget plot]
table [row sep=\\]{%
14706	77.7821868835 \\
};
\addplot [semithick, color0, mark=*, mark options={solid}, forget plot]
table [row sep=\\]{%
40306	71.6701873161 \\
};
\addplot [semithick, color0, mark=*, mark options={solid}, forget plot]
table [row sep=\\]{%
78706	69.731599247 \\
};
\addplot [semithick, color0, mark=*, mark options={solid}, forget plot]
table [row sep=\\]{%
2402	45.0585057479 \\
};
\addplot [semithick, color0, mark=*, mark options={solid}, forget plot]
table [row sep=\\]{%
19810	70.2822457563 \\
};
\addplot [semithick, color0, mark=*, mark options={solid}, forget plot]
table [row sep=\\]{%
54626	67.5552148422 \\
};
\addplot [semithick, color0, mark=*, mark options={solid}, forget plot]
table [row sep=\\]{%
106850	64.4608856116 \\
};
\addplot [semithick, color0, mark=*, mark options={solid}, forget plot]
table [row sep=\\]{%
4642	56.0325446817 \\
};
\addplot [semithick, color0, mark=*, mark options={solid}, forget plot]
table [row sep=\\]{%
38434	72.2025643146 \\
};
\addplot [semithick, color0, mark=*, mark options={solid}, forget plot]
table [row sep=\\]{%
106018	69.9391154125 \\
};
\addplot [semithick, color0, mark=*, mark options={solid}, forget plot]
table [row sep=\\]{%
207394	67.8389944747 \\
};
\addplot [semithick, color0, mark=*, mark options={solid}, forget plot]
table [row sep=\\]{%
6882	51.7281468612 \\
};
\addplot [semithick, color0, mark=*, mark options={solid}, forget plot]
table [row sep=\\]{%
57058	70.4458515267 \\
};
\addplot [semithick, color0, mark=*, mark options={solid}, forget plot]
table [row sep=\\]{%
157410	71.1520880482 \\
};
\addplot [semithick, color0, mark=*, mark options={solid}, forget plot]
table [row sep=\\]{%
307938	68.6509241651 \\
};
\addplot [semithick, color0, mark=*, mark size=2, mark options={solid}, forget plot]
table [row sep=\\]{%
32	27.5409578822 \\
};
\addplot [semithick, color0, mark=*, mark size=2, mark options={solid,draw=black}, forget plot]
table [row sep=\\]{%
160	73.6680486383 \\
};
\addplot [semithick, color0, mark=*, mark size=2, mark options={solid}, forget plot]
table [row sep=\\]{%
416	59.9266677766 \\
};
\addplot [semithick, color0, mark=*, mark size=2, mark options={solid}, forget plot]
table [row sep=\\]{%
800	57.6773904145 \\
};
\addplot [semithick, color0, mark=*, mark size=2, mark options={solid}, forget plot]
table [row sep=\\]{%
52	-0.1576405369 \\
};
\addplot [semithick, color0, mark=*, mark size=2, mark options={solid}, forget plot]
table [row sep=\\]{%
244	58.5909080674 \\
};
\addplot [semithick, color0, mark=*, mark size=2, mark options={solid}, forget plot]
table [row sep=\\]{%
628	58.1844080994 \\
};
\addplot [semithick, color0, mark=*, mark size=2, mark options={solid}, forget plot]
table [row sep=\\]{%
1204	57.3168772462 \\
};
\addplot [semithick, color0, mark=*, mark size=2, mark options={solid}, forget plot]
table [row sep=\\]{%
72	-0 \\
};
\addplot [semithick, color0, mark=*, mark size=2, mark options={solid}, forget plot]
table [row sep=\\]{%
328	55.7072683787 \\
};
\addplot [semithick, color0, mark=*, mark size=2, mark options={solid}, forget plot]
table [row sep=\\]{%
840	59.4853093099 \\
};
\addplot [semithick, color0, mark=*, mark size=2, mark options={solid}, forget plot]
table [row sep=\\]{%
1608	59.8816655165 \\
};
\addplot [semithick, color0, mark=*, mark size=2, mark options={solid}, forget plot]
table [row sep=\\]{%
218	66.2288694122 \\
};
\addplot [semithick, color0, mark=*, mark size=2, mark options={solid}, forget plot]
table [row sep=\\]{%
1498	66.4430816339 \\
};
\addplot [semithick, color0, mark=*, mark size=2, mark options={solid}, forget plot]
table [row sep=\\]{%
4058	62.7552543974 \\
};
\addplot [semithick, color0, mark=*, mark size=2, mark options={solid}, forget plot]
table [row sep=\\]{%
7898	62.0441027812 \\
};
\addplot [semithick, color0, mark=*, mark size=2, mark options={solid}, forget plot]
table [row sep=\\]{%
394	56.8334041743 \\
};
\addplot [semithick, color0, mark=*, mark size=2, mark options={solid}, forget plot]
table [row sep=\\]{%
2698	65.1447815661 \\
};
\addplot [semithick, color0, mark=*, mark size=2, mark options={solid}, forget plot]
table [row sep=\\]{%
7306	66.3018892378 \\
};
\addplot [semithick, color0, mark=*, mark size=2, mark options={solid}, forget plot]
table [row sep=\\]{%
14218	62.7459427801 \\
};
\addplot [semithick, color0, mark=*, mark size=2, mark options={solid}, forget plot]
table [row sep=\\]{%
570	54.4928935087 \\
};
\addplot [semithick, color0, mark=*, mark size=2, mark options={solid,draw=black}, forget plot]
table [row sep=\\]{%
3898	73.5985328336 \\
};
\addplot [semithick, color0, mark=*, mark size=2, mark options={solid}, forget plot]
table [row sep=\\]{%
10554	68.8800019494 \\
};
\addplot [semithick, color0, mark=*, mark size=2, mark options={solid}, forget plot]
table [row sep=\\]{%
20538	65.1940938149 \\
};
\addplot [semithick, color0, mark=*, mark size=2, mark options={solid}, forget plot]
table [row sep=\\]{%
690	61.6360151458 \\
};
\addplot [semithick, color0, mark=*, mark size=2, mark options={solid,draw=black}, forget plot]
table [row sep=\\]{%
5298	74.5742588344 \\
};
\addplot [semithick, color0, mark=*, mark size=2, mark options={solid}, forget plot]
table [row sep=\\]{%
14514	66.3420960590 \\
};
\addplot [semithick, color0, mark=*, mark size=2, mark options={solid}, forget plot]
table [row sep=\\]{%
28338	63.5896587443 \\
};
\addplot [semithick, color0, mark=*, mark size=2, mark options={solid}, forget plot]
table [row sep=\\]{%
1298	66.8702718026 \\
};
\addplot [semithick, color0, mark=*, mark size=2, mark options={solid,draw=black}, forget plot]
table [row sep=\\]{%
10002	77.4738937999 \\
};
\addplot [semithick, color0, mark=*, mark size=2, mark options={solid}, forget plot]
table [row sep=\\]{%
27410	63.0885652024 \\
};
\addplot [semithick, color0, mark=*, mark size=2, mark options={solid}, forget plot]
table [row sep=\\]{%
53522	65.6939642661 \\
};
\addplot [semithick, color0, mark=*, mark size=2, mark options={solid}, forget plot]
table [row sep=\\]{%
1906	57.2404149553 \\
};
\addplot [semithick, color0, mark=*, mark size=2, mark options={solid,draw=black}, forget plot]
table [row sep=\\]{%
14706	77.7821868835 \\
};
\addplot [semithick, color0, mark=*, mark size=2, mark options={solid}, forget plot]
table [row sep=\\]{%
40306	71.6701873161 \\
};
\addplot [semithick, color0, mark=*, mark size=2, mark options={solid}, forget plot]
table [row sep=\\]{%
78706	69.731599247 \\
};
\addplot [semithick, color0, mark=*, mark size=2, mark options={solid}, forget plot]
table [row sep=\\]{%
2402	45.0585057479 \\
};
\addplot [semithick, color0, mark=*, mark size=2, mark options={solid}, forget plot]
table [row sep=\\]{%
19810	70.2822457563 \\
};
\addplot [semithick, color0, mark=*, mark size=2, mark options={solid}, forget plot]
table [row sep=\\]{%
54626	67.5552148422 \\
};
\addplot [semithick, color0, mark=*, mark size=2, mark options={solid}, forget plot]
table [row sep=\\]{%
106850	64.4608856116 \\
};
\addplot [semithick, color0, mark=*, mark size=2, mark options={solid}, forget plot]
table [row sep=\\]{%
4642	56.0325446817 \\
};
\addplot [semithick, color0, mark=*, mark size=2, mark options={solid,draw=black}, forget plot]
table [row sep=\\]{%
38434	72.2025643146 \\
};
\addplot [semithick, color0, mark=*, mark size=2, mark options={solid}, forget plot]
table [row sep=\\]{%
106018	69.9391154125 \\
};
\addplot [semithick, color0, mark=*, mark size=2, mark options={solid}, forget plot]
table [row sep=\\]{%
207394	67.8389944747 \\
};
\addplot [semithick, color0, mark=*, mark size=2, mark options={solid}, forget plot]
table [row sep=\\]{%
6882	51.7281468612 \\
};
\addplot [semithick, color0, mark=*, mark size=2, mark options={solid}, forget plot]
table [row sep=\\]{%
57058	70.4458515267 \\
};
\addplot [semithick, color0, mark=*, mark size=2, mark options={solid}, forget plot]
table [row sep=\\]{%
157410	71.1520880482 \\
};
\addplot [semithick, color0, mark=*, mark size=2, mark options={solid}, forget plot]
table [row sep=\\]{%
307938	68.6509241651 \\
};

\draw[] (axis cs:160,73.6680486383) -- (axis cs:160,73.6680486383);
\node [node on layer=front] at (axis cs:160,73.6680486383)[
  scale=0.8,
text=black,
rotate=0.0,
shift={(-0.6em,-0.4em)}
]{\bfseries A};
\draw[] (axis cs:3898,73.5985328336) -- (axis cs:3898,73.5985328336);
\node [node on layer=front] at (axis cs:3898,73.5985328336)[
  scale=0.8,
text=black,
rotate=0.0,
shift={(-0.6em,0.6em)}
]{\bfseries B};
\draw[] (axis cs:5298,74.5742588344) -- (axis cs:5298,74.5742588344);
\node [node on layer=front] at (axis cs:5298,74.5742588344)[
  scale=0.8,
text=black,
rotate=0.0,
shift={(0.5em,0.5em)}
]{\bfseries C};
\draw[] (axis cs:10002,77.4738937999) -- (axis cs:10002,77.4738937999);
\node [node on layer=front] at (axis cs:10002,77.4738937999)[
  scale=0.8,
text=black,
rotate=0.0,
shift={(-0.5em,0.6em)}
]{\bfseries D};
\draw[] (axis cs:14706,77.7821868835) -- (axis cs:14706,77.7821868835);
\node [node on layer=front] at (axis cs:14706,77.7821868835)[
  scale=0.8,
text=black,
rotate=0.0,
shift={(0.6em,0.6em)}
]{\bfseries E};
\draw[] (axis cs:38434,72.2025643146) -- (axis cs:38434,72.2025643146);
\node [node on layer=front] at (axis cs:38434,72.2025643146)[
  scale=0.8,
text=black,
rotate=0.0,
shift={(-0.6em,0.6em)}
]{\bfseries F};
\end{axis}

\begin{axis}[
axis background,
axis line style={opacity=0},
every major grid/.style={opacity=0},
axis y line=right,
tick align=outside,
x grid style={line width = 0pt},
xmajorgrids,
xmin=20.2294610720991, xmax=487112.136348054,
xmode=log,
xtick pos=left,
y grid style={line width = 0pt},
ylabel={\textcolor{color1}{EVM}},
ymajorgrids,
ymin=-0.04999999999, ymax=1.04999999979,
ytick pos=right,
yticklabel style=color1,
ytick={-0.2,0,0.2,0.4,0.6,0.8,1,1.2},
yticklabels={−0.2,0.0,0.2,0.4,0.6,0.8,1.0,1.2},
xticklabels={},
xlabel={}
]
\addplot [semithick, color1, mark=*, mark options={solid}, forget plot]
table [row sep=\\]{%
32	0.9826458211 \\
};
\addplot [semithick, color1, mark=*, mark options={solid}, forget plot]
table [row sep=\\]{%
160	0.9009347409 \\
};
\addplot [semithick, color1, mark=*, mark options={solid}, forget plot]
table [row sep=\\]{%
416	0.6730611048 \\
};
\addplot [semithick, color1, mark=*, mark options={solid}, forget plot]
table [row sep=\\]{%
800	0.2475016456 \\
};
\addplot [semithick, color1, mark=*, mark options={solid}, forget plot]
table [row sep=\\]{%
52	0.9999962951 \\
};
\addplot [semithick, color1, mark=*, mark options={solid}, forget plot]
table [row sep=\\]{%
244	0.6920939888 \\
};
\addplot [semithick, color1, mark=*, mark options={solid}, forget plot]
table [row sep=\\]{%
628	0.5000736714 \\
};
\addplot [semithick, color1, mark=*, mark options={solid}, forget plot]
table [row sep=\\]{%
1204	0.6021769587 \\
};
\addplot [semithick, color1, mark=*, mark options={solid}, forget plot]
table [row sep=\\]{%
72	0.9999999998 \\
};
\addplot [semithick, color1, mark=*, mark options={solid}, forget plot]
table [row sep=\\]{%
328	0.41026755 \\
};
\addplot [semithick, color1, mark=*, mark options={solid}, forget plot]
table [row sep=\\]{%
840	0.4724356965 \\
};
\addplot [semithick, color1, mark=*, mark options={solid}, forget plot]
table [row sep=\\]{%
1608	0.2547774838 \\
};
\addplot [semithick, color1, mark=*, mark options={solid}, forget plot]
table [row sep=\\]{%
218	0.6088722732 \\
};
\addplot [semithick, color1, mark=*, mark options={solid}, forget plot]
table [row sep=\\]{%
1498	0.1786671015 \\
};
\addplot [semithick, color1, mark=*, mark options={solid}, forget plot]
table [row sep=\\]{%
4058	0.1648493597 \\
};
\addplot [semithick, color1, mark=*, mark options={solid}, forget plot]
table [row sep=\\]{%
7898	0.257943521 \\
};
\addplot [semithick, color1, mark=*, mark options={solid}, forget plot]
table [row sep=\\]{%
394	0.8473671589 \\
};
\addplot [semithick, color1, mark=*, mark options={solid}, forget plot]
table [row sep=\\]{%
2698	0.2909147153 \\
};
\addplot [semithick, color1, mark=*, mark options={solid}, forget plot]
table [row sep=\\]{%
7306	0.602058275 \\
};
\addplot [semithick, color1, mark=*, mark options={solid}, forget plot]
table [row sep=\\]{%
14218	0.3424733354 \\
};
\addplot [semithick, color1, mark=*, mark options={solid}, forget plot]
table [row sep=\\]{%
570	0.7719184933 \\
};
\addplot [semithick, color1, mark=*, mark options={solid}, forget plot]
table [row sep=\\]{%
3898	0.4017015079 \\
};
\addplot [semithick, color1, mark=*, mark options={solid}, forget plot]
table [row sep=\\]{%
10554	0.3269122533 \\
};
\addplot [semithick, color1, mark=*, mark options={solid}, forget plot]
table [row sep=\\]{%
20538	0.1596120995 \\
};
\addplot [semithick, color1, mark=*, mark options={solid}, forget plot]
table [row sep=\\]{%
690	0.6151190557 \\
};
\addplot [semithick, color1, mark=*, mark options={solid}, forget plot]
table [row sep=\\]{%
5298	0.2956641572 \\
};
\addplot [semithick, color1, mark=*, mark options={solid}, forget plot]
table [row sep=\\]{%
14514	0.1446697931 \\
};
\addplot [semithick, color1, mark=*, mark options={solid}, forget plot]
table [row sep=\\]{%
28338	0.1488966187 \\
};
\addplot [semithick, color1, mark=*, mark options={solid}, forget plot]
table [row sep=\\]{%
1298	0.7736031308 \\
};
\addplot [semithick, color1, mark=*, mark options={solid}, forget plot]
table [row sep=\\]{%
10002	0.5765175565 \\
};
\addplot [semithick, color1, mark=*, mark options={solid}, forget plot]
table [row sep=\\]{%
27410	0.3271419173 \\
};
\addplot [semithick, color1, mark=*, mark options={solid}, forget plot]
table [row sep=\\]{%
53522	0.1972290333 \\
};
\addplot [semithick, color1, mark=*, mark options={solid}, forget plot]
table [row sep=\\]{%
1906	0.4481231811 \\
};
\addplot [semithick, color1, mark=*, mark options={solid}, forget plot]
table [row sep=\\]{%
14706	0.4690600331 \\
};
\addplot [semithick, color1, mark=*, mark options={solid}, forget plot]
table [row sep=\\]{%
40306	0.2194023333 \\
};
\addplot [semithick, color1, mark=*, mark options={solid}, forget plot]
table [row sep=\\]{%
78706	0.1596266563 \\
};
\addplot [semithick, color1, mark=*, mark options={solid}, forget plot]
table [row sep=\\]{%
2402	0.7941948278 \\
};
\addplot [semithick, color1, mark=*, mark options={solid}, forget plot]
table [row sep=\\]{%
19810	0.1465261495 \\
};
\addplot [semithick, color1, mark=*, mark options={solid}, forget plot]
table [row sep=\\]{%
54626	0.1106476408 \\
};
\addplot [semithick, color1, mark=*, mark options={solid}, forget plot]
table [row sep=\\]{%
106850	0.1154832234 \\
};
\addplot [semithick, color1, mark=*, mark options={solid}, forget plot]
table [row sep=\\]{%
4642	0.7305572872 \\
};
\addplot [semithick, color1, mark=*, mark options={solid}, forget plot]
table [row sep=\\]{%
38434	0.5144240469 \\
};
\addplot [semithick, color1, mark=*, mark options={solid}, forget plot]
table [row sep=\\]{%
106018	0.4705189172 \\
};
\addplot [semithick, color1, mark=*, mark options={solid}, forget plot]
table [row sep=\\]{%
207394	0.2705358505 \\
};
\addplot [semithick, color1, mark=*, mark options={solid}, forget plot]
table [row sep=\\]{%
6882	0.7366814099 \\
};
\addplot [semithick, color1, mark=*, mark options={solid}, forget plot]
table [row sep=\\]{%
57058	0.386727402 \\
};
\addplot [semithick, color1, mark=*, mark options={solid}, forget plot]
table [row sep=\\]{%
157410	0.295158557 \\
};
\addplot [semithick, color1, mark=*, mark options={solid}, forget plot]
table [row sep=\\]{%
307938	0.1412663061 \\
};
\addplot [semithick, color1, mark=*, mark size=2, mark options={solid}, forget plot]
table [row sep=\\]{%
32	0.9826458211 \\
};
\addplot [semithick, color1, mark=*, mark size=2, mark options={solid,draw=black}, forget plot]
table [row sep=\\]{%
160	0.9009347409 \\
};
\addplot [semithick, color1, mark=*, mark size=2, mark options={solid}, forget plot]
table [row sep=\\]{%
416	0.6730611048 \\
};
\addplot [semithick, color1, mark=*, mark size=2, mark options={solid}, forget plot]
table [row sep=\\]{%
800	0.2475016456 \\
};
\addplot [semithick, color1, mark=*, mark size=2, mark options={solid}, forget plot]
table [row sep=\\]{%
52	0.9999962951 \\
};
\addplot [semithick, color1, mark=*, mark size=2, mark options={solid}, forget plot]
table [row sep=\\]{%
244	0.6920939888 \\
};
\addplot [semithick, color1, mark=*, mark size=2, mark options={solid}, forget plot]
table [row sep=\\]{%
628	0.5000736714 \\
};
\addplot [semithick, color1, mark=*, mark size=2, mark options={solid}, forget plot]
table [row sep=\\]{%
1204	0.6021769587 \\
};
\addplot [semithick, color1, mark=*, mark size=2, mark options={solid}, forget plot]
table [row sep=\\]{%
72	0.9999999998 \\
};
\addplot [semithick, color1, mark=*, mark size=2, mark options={solid}, forget plot]
table [row sep=\\]{%
328	0.41026755 \\
};
\addplot [semithick, color1, mark=*, mark size=2, mark options={solid}, forget plot]
table [row sep=\\]{%
840	0.4724356965 \\
};
\addplot [semithick, color1, mark=*, mark size=2, mark options={solid}, forget plot]
table [row sep=\\]{%
1608	0.2547774838 \\
};
\addplot [semithick, color1, mark=*, mark size=2, mark options={solid}, forget plot]
table [row sep=\\]{%
218	0.6088722732 \\
};
\addplot [semithick, color1, mark=*, mark size=2, mark options={solid}, forget plot]
table [row sep=\\]{%
1498	0.1786671015 \\
};
\addplot [semithick, color1, mark=*, mark size=2, mark options={solid}, forget plot]
table [row sep=\\]{%
4058	0.1648493597 \\
};
\addplot [semithick, color1, mark=*, mark size=2, mark options={solid}, forget plot]
table [row sep=\\]{%
7898	0.257943521 \\
};
\addplot [semithick, color1, mark=*, mark size=2, mark options={solid}, forget plot]
table [row sep=\\]{%
394	0.8473671589 \\
};
\addplot [semithick, color1, mark=*, mark size=2, mark options={solid}, forget plot]
table [row sep=\\]{%
2698	0.2909147153 \\
};
\addplot [semithick, color1, mark=*, mark size=2, mark options={solid}, forget plot]
table [row sep=\\]{%
7306	0.602058275 \\
};
\addplot [semithick, color1, mark=*, mark size=2, mark options={solid}, forget plot]
table [row sep=\\]{%
14218	0.3424733354 \\
};
\addplot [semithick, color1, mark=*, mark size=2, mark options={solid}, forget plot]
table [row sep=\\]{%
570	0.7719184933 \\
};
\addplot [semithick, color1, mark=*, mark size=2, mark options={solid,draw=black}, forget plot]
table [row sep=\\]{%
3898	0.4017015079 \\
};
\addplot [semithick, color1, mark=*, mark size=2, mark options={solid}, forget plot]
table [row sep=\\]{%
10554	0.3269122533 \\
};
\addplot [semithick, color1, mark=*, mark size=2, mark options={solid}, forget plot]
table [row sep=\\]{%
20538	0.1596120995 \\
};
\addplot [semithick, color1, mark=*, mark size=2, mark options={solid}, forget plot]
table [row sep=\\]{%
690	0.6151190557 \\
};
\addplot [semithick, color1, mark=*, mark size=2, mark options={solid,draw=black}, forget plot]
table [row sep=\\]{%
5298	0.2956641572 \\
};
\addplot [semithick, color1, mark=*, mark size=2, mark options={solid}, forget plot]
table [row sep=\\]{%
14514	0.1446697931 \\
};
\addplot [semithick, color1, mark=*, mark size=2, mark options={solid}, forget plot]
table [row sep=\\]{%
28338	0.1488966187 \\
};
\addplot [semithick, color1, mark=*, mark size=2, mark options={solid}, forget plot]
table [row sep=\\]{%
1298	0.7736031308 \\
};
\addplot [semithick, color1, mark=*, mark size=2, mark options={solid,draw=black}, forget plot]
table [row sep=\\]{%
10002	0.5765175565 \\
};
\addplot [semithick, color1, mark=*, mark size=2, mark options={solid}, forget plot]
table [row sep=\\]{%
27410	0.3271419173 \\
};
\addplot [semithick, color1, mark=*, mark size=2, mark options={solid}, forget plot]
table [row sep=\\]{%
53522	0.1972290333 \\
};
\addplot [semithick, color1, mark=*, mark size=2, mark options={solid}, forget plot]
table [row sep=\\]{%
1906	0.4481231811 \\
};
\addplot [semithick, color1, mark=*, mark size=2, mark options={solid,draw=black}, forget plot]
table [row sep=\\]{%
14706	0.4690600331 \\
};
\addplot [semithick, color1, mark=*, mark size=2, mark options={solid}, forget plot]
table [row sep=\\]{%
40306	0.2194023333 \\
};
\addplot [semithick, color1, mark=*, mark size=2, mark options={solid}, forget plot]
table [row sep=\\]{%
78706	0.1596266563 \\
};
\addplot [semithick, color1, mark=*, mark size=2, mark options={solid}, forget plot]
table [row sep=\\]{%
2402	0.7941948278 \\
};
\addplot [semithick, color1, mark=*, mark size=2, mark options={solid}, forget plot]
table [row sep=\\]{%
19810	0.1465261495 \\
};
\addplot [semithick, color1, mark=*, mark size=2, mark options={solid}, forget plot]
table [row sep=\\]{%
54626	0.1106476408 \\
};
\addplot [semithick, color1, mark=*, mark size=2, mark options={solid}, forget plot]
table [row sep=\\]{%
106850	0.1154832234 \\
};
\addplot [semithick, color1, mark=*, mark size=2, mark options={solid}, forget plot]
table [row sep=\\]{%
4642	0.7305572872 \\
};
\addplot [semithick, color1, mark=*, mark size=2, mark options={solid,draw=black}, forget plot]
table [row sep=\\]{%
38434	0.5144240469 \\
};
\addplot [semithick, color1, mark=*, mark size=2, mark options={solid}, forget plot]
table [row sep=\\]{%
106018	0.4705189172 \\
};
\addplot [semithick, color1, mark=*, mark size=2, mark options={solid}, forget plot]
table [row sep=\\]{%
207394	0.2705358505 \\
};
\addplot [semithick, color1, mark=*, mark size=2, mark options={solid}, forget plot]
table [row sep=\\]{%
6882	0.7366814099 \\
};
\addplot [semithick, color1, mark=*, mark size=2, mark options={solid}, forget plot]
table [row sep=\\]{%
57058	0.386727402 \\
};
\addplot [semithick, color1, mark=*, mark size=2, mark options={solid}, forget plot]
table [row sep=\\]{%
157410	0.295158557 \\
};
\addplot [semithick, color1, mark=*, mark size=2, mark options={solid}, forget plot]
table [row sep=\\]{%
307938	0.1412663061 \\
};

\draw[] (axis cs:160,0.9009347409) -- (axis cs:160,0.9009347409);
\node [node on layer=front] at (axis cs:160,0.9009347409)[
  scale=0.8,
  text=black,
  rotate=0.0,
  shift={(0.6em,0.6em)}
]{\bfseries A};
\draw[] (axis cs:3898,0.4017015079) -- (axis cs:3898,0.4017015079);
\node [node on layer=front] at (axis cs:3898,0.4017015079)[
  scale=0.8,
text=black,
rotate=0.0,
shift={(-0.6em,0.6em)}
]{\bfseries B};
\draw[] (axis cs:5298,0.2956641572) -- (axis cs:5298,0.2956641572);
\node [node on layer=front] at (axis cs:5298,0.2956641572)[
  scale=0.8,
text=black,
rotate=0.0,
shift={(0.6em,0.6em)}
]{\bfseries C};
\draw[] (axis cs:10002,0.5765175565) -- (axis cs:10002,0.5765175565);
\node [node on layer=front] at (axis cs:10002,0.5765175565)[
  scale=0.8,
text=black,
rotate=0.0,
shift={(0.6em,0.6em)}
]{\bfseries D};
\draw[] (axis cs:14706,0.4690600331) -- (axis cs:14706,0.4690600331);
\node [node on layer=front] at (axis cs:14706,0.4690600331)[
  scale=0.8,
text=black,
rotate=0.0,
shift={(-0.7em,-0.4em)}
]{\bfseries E};
\draw[] (axis cs:38434,0.5144240469) -- (axis cs:38434,0.5144240469);
\node [node on layer=front] at (axis cs:38434,0.5144240469)[
  scale=0.8,
text=black,
rotate=0.0,
shift={(0.6em,0.6em)}
]{\bfseries F};
\end{axis}

\begin{axis}[
set layers,axis background,
axis y line=right,
tick align=outside,
x grid style={line width = 0pt},
xmajorgrids,
xmin=20.2294610720991, xmax=487112.136348054,
xmode=log,
xtick pos=left,
y grid style={line width = 0pt},
ymajorgrids,
ymin=-0.04999999999, ymax=1.04999999979,
ytick pos=right,
ytick={-0.2,0,0.2,0.4,0.6,0.8,1,1.2},
yticklabels={},
xticklabels={},
xlabel={}
]
\end{axis}

\end{tikzpicture}

%% file: results_comparison.tex
\subsection{Comparative Analysis with Other Techniques}

The algorithms described in Section~\ref{subsec:setupcomp} were implemented in the simulation framework and evaluated using performance metrics from Section~\ref{subsec:perfmeas}. The results were statistically compared with the best-performing CNN-based method (RD RIS Model D) from Section~\ref{sec:performance-cnn}.

\begin{figure}[tb]
	\centering
	\resizebox {0.9\columnwidth} {!} {
		\input{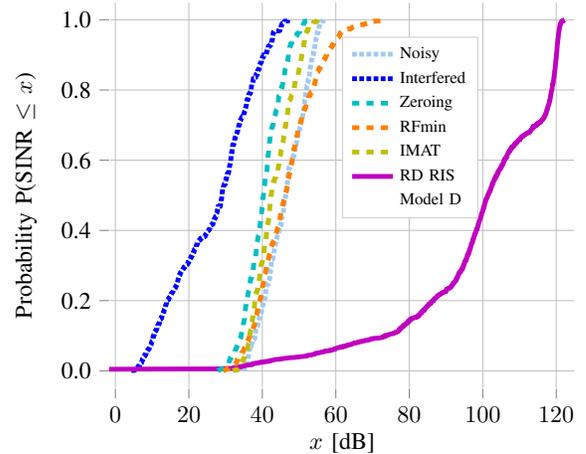}
	}
	\caption{CDF comparison of RD SINR with other techniques.}
	\label{fig:cdf-rd-sinr}
\end{figure}

Fig.~\ref{fig:cdf-rd-sinr} shows the range-Doppler SINR performance of the selected techniques. Zeroing already considerably increases the SINR, since it removes interference completely (due to the perfect detection assumption), although it also removes parts of the object signal. IMAT is a natural improvement of zeroing, while ramp filtering can achieve an even larger noise suppression due to the principle of its non-linear operation. RD RIS denoising results in a superior average SINR compared to all other approaches. In fact, it appears to implicitly detect object peaks, enabling it to maximally attenuate the surrounding noise. It is only in severely interfered scenarios that object peaks are not recognized, and thus suppressed. This is shown by the long tail of the SINR CDF.

\begin{figure}[tb]
	\centering
	\resizebox {0.9\columnwidth} {!} {
		\input{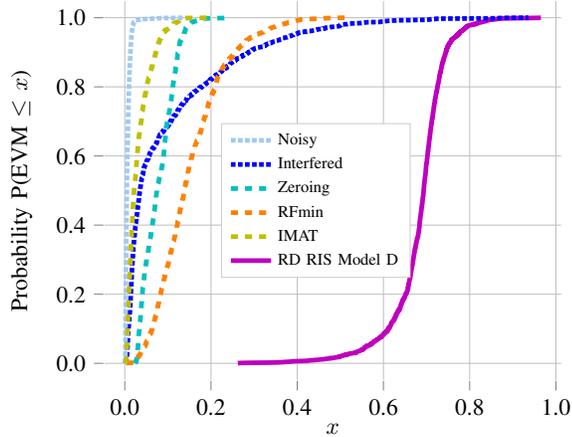}
	}
	\caption{CDF comparison of RD EVM with other techniques.}
	\label{fig:cdf-rd-evm}
\end{figure}

Fig.~\ref{fig:cdf-rd-evm} shows the corresponding EVM performance. Due to its non-linear nature, ramp filtering performs the worst among conventional methods. IMAT performs the best, reducing the bias in object peak values introduced by zeroing. RD RIS denoising on the other hand yields much higher EVMs. This indicates that object value preservation cannot be reliably guaranteed by such a denoising method.

\begin{table}
	\centering
	\caption{Performance comparison with state-of-the-art interference mitigation methods.}
	\begin{tabular}{l >{\bfseries}r r >{\bfseries}r r}
		Signal & SINR (RD) & EVM (RD) \\
		\hline
		Noisy & 46.16 & 0.01 \\
		Interfered & 26.67 & 0.10 \\
		Zeroing & 40.42 & 0.08 \\
		Ramp filtering  & 46.50 & 0.15 \\
		IMAT & 43.20 & 0.03 \\
		RD RIS Model D & 98.84 & 0.78 \\
	\end{tabular}
	\label{tab:performance-comparison}
\end{table}

Table~\ref{tab:performance-comparison} shows the mean values of performance metrics over the simulated test scenarios. For the conventional methods, ramp filtering achieves the highest SINR at the cost of an elevated EVM, while IMAT can improve on the performance of zeroing and lower the EVM. RD RIS denoising outperforms all the conventional methods in terms of SINR, while the average EVM is considerably higher.

\begin{figure}
	\centering
	\resizebox {0.9\columnwidth} {!} {
		\input{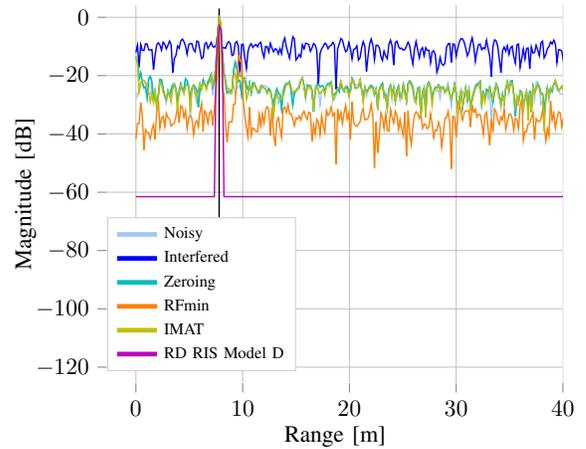}
	}
	\caption{Range cut at velocity $v=5.5\textrm{m/s}$, shown up to $40\textrm{m}$. The object is located at a distance $d=7.9 \textrm{m}$ as indicated by the vertical black marker.}
	\label{fig:range-cut}
\end{figure}

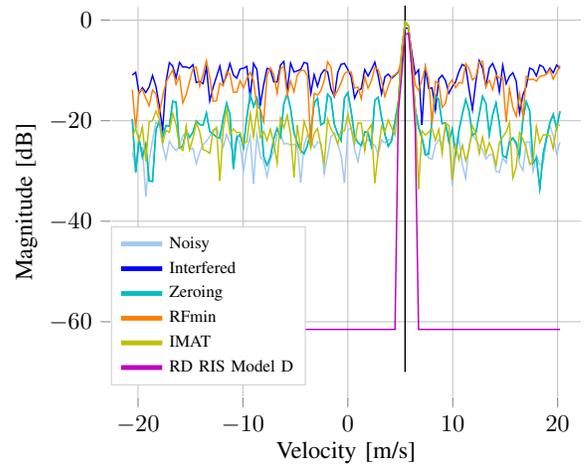
\begin{figure}
	\centering
	\resizebox {0.9\columnwidth} {!} {
		\input{velocity_cut.tex}
	}
	\caption{Velocity cut at distance $d=7.9 \textrm{m}$. The object is located at a velocity $v=5.5\textrm{m/s}$ as indicated by the vertical black marker.}
	\label{fig:velocity-cut}
\end{figure}

In order to illustrate the effects of the different mitigation methods, Figs.~\ref{fig:range-cut} and~\ref{fig:velocity-cut} show range- and velocity cuts of a RD map. The magnitude-normalized RD map of the first receive channel is plotted as a log-magnitude spectrum at a distance $d=7.9 \textrm{m}$ and velocity $v=5.5 \textrm{m/s}$ respectively. It can be noted that ramp filtering strongly suppresses noise, though mainly on the range axis. Zeroing and IMAT have almost the same effect for the visualized scenario. However, the previous statistical analysis shows that IMAT is superior, especially regarding object peak value preservation. The RD RIS model enhances object peaks, while it strongly reduces the noise floor to a \emph{constant} level. This reassures our presumption that the CNN-based denoising has an implicit thresholding effect.

In summary, CNN-based methods have superior noise and interference suppression capabilities compared to conventional algorithms. However, some exhibit a considerably high EVM which may lead to distortions in object peak values. Such distortions may have negative effects on further radar processing, such as on angular estimation or object classification. Due to these properties, the CNN-denoised signal is very well suited for object detection on the RD map while further processing can alternatively be performed using the non-mitigated object peak values.

%% file: velocity_cut.tex
\begin{tikzpicture}

\definecolor{color1}{rgb}{0.75,0,0.75}
\definecolor{color2}{rgb}{0.75,0.75,0}
\definecolor{color0}{rgb}{0,0.75,0.75}
\definecolor{babyblueeyes}{rgb}{0.63, 0.79, 0.95}

\begin{axis}[
	axis line style={white},
	legend cell align={left},
	legend entries={{Noisy},{Interfered},{Zeroing},{RFmin},{IMAT},{RD RIS Model D}},
	legend style={at={(0,0.03)}, anchor=south west, draw=white!80.0!black, font=\scriptsize},
	tick align=outside,
	tick pos=left,
	xlabel={Velocity [m/s]},
	xmajorgrids,
	xmin=-22.5836750356764, xmax=22.2626562007698,
	ylabel={Magnitude [dB]},
	ymajorgrids,
	ymin=-74.957540435191, ymax=2.74296939224381
	]
	
	\addlegendimage{solid, no markers, babyblueeyes, line width=2pt}
	\addlegendimage{solid, no markers, blue, line width=2pt}
	\addlegendimage{solid, no markers, color0, line width=2pt}
	\addlegendimage{solid, no markers, orange, line width=2pt}
	\addlegendimage{solid, no markers, color2, line width=2pt}
	\addlegendimage{solid, no markers, color1, line width=2pt}
	
	\addplot [semithick, black, forget plot]
	table [row sep=\\]{%
		5.4573201934115	-70 \\
		5.4573201934115	3 \\
	};

\addplot [semithick, babyblueeyes]
table [row sep=\\]{%
	-20.5452054340198	-26.3374029452968 \\
	-20.2241865991132	-29.0789380401773 \\
	-19.9031677642066	-25.4450203039185 \\
	-19.5821489293001	-25.3322168087373 \\
	-19.2611300943935	-34.9939148435476 \\
	-18.940111259487	-27.8688298190674 \\
	-18.6190924245804	-23.8659733709701 \\
	-18.2980735896738	-20.3191901412004 \\
	-17.9770547547673	-20.6714107591678 \\
	-17.6560359198607	-24.9552039837842 \\
	-17.3350170849542	-27.6274052874654 \\
	-17.0139982500476	-24.0855013026055 \\
	-16.6929794151411	-25.1891837482557 \\
	-16.3719605802345	-25.9466320273137 \\
	-16.0509417453279	-25.0968647834561 \\
	-15.7299229104214	-23.0003377398461 \\
	-15.4089040755148	-22.3281832529041 \\
	-15.0878852406083	-25.6569726546775 \\
	-14.7668664057017	-25.4928225101076 \\
	-14.4458475707951	-28.5552292457271 \\
	-14.1248287358886	-26.2787476165276 \\
	-13.803809900982	-24.7455711218063 \\
	-13.4827910660755	-24.1079324700041 \\
	-13.1617722311689	-23.1238436666547 \\
	-12.8407533962623	-22.265189834726 \\
	-12.5197345613558	-21.5098913571923 \\
	-12.1987157264492	-23.1389379847807 \\
	-11.8776968915427	-25.1971174316548 \\
	-11.5566780566361	-22.7451907296378 \\
	-11.2356592217296	-24.5794111442485 \\
	-10.914640386823	-26.4318530704113 \\
	-10.5936215519164	-22.9712591459797 \\
	-10.2726027170099	-24.6587698650569 \\
	-9.95158388210332	-30.004815017163 \\
	-9.63056504719676	-25.7455942263018 \\
	-9.3095462122902	-23.0536888266171 \\
	-8.98852737738364	-32.5282583910095 \\
	-8.66750854247709	-22.8653192187207 \\
	-8.34648970757053	-27.186283735224 \\
	-8.02547087266397	-28.0174538569003 \\
	-7.70445203775741	-31.1127941843649 \\
	-7.38343320285085	-27.782873059493 \\
	-7.06241436794429	-26.0920470586613 \\
	-6.74139553303773	-22.7554462037223 \\
	-6.42037669813117	-22.3272395782092 \\
	-6.09935786322461	-24.6215599884011 \\
	-5.77833902831806	-24.5282557509142 \\
	-5.4573201934115	-24.8561847423158 \\
	-5.13630135850494	-24.7113079050669 \\
	-4.81528252359838	-24.9962238590802 \\
	-4.49426368869182	-32.3424789070825 \\
	-4.17324485378526	-24.8478864883905 \\
	-3.8522260188787	-22.8624518433243 \\
	-3.53120718397215	-23.7622950951649 \\
	-3.21018834906559	-25.3815677540082 \\
	-2.88916951415903	-24.2460166823269 \\
	-2.56815067925247	-24.5176806255613 \\
	-2.24713184434591	-30.5416966253476 \\
	-1.92611300943935	-26.8338605558645 \\
	-1.60509417453279	-23.1802074697799 \\
	-1.28407533962623	-22.4781168293479 \\
	-0.963056504719676	-24.6368954748621 \\
	-0.642037669813117	-26.9099648896539 \\
	-0.321018834906559	-32.345999270994 \\
	0	-21.7542149179309 \\
	0.321018834906559	-21.5150654671811 \\
	0.642037669813117	-25.7208575588469 \\
	0.963056504719676	-28.1251968882898 \\
	1.28407533962623	-24.4242133520265 \\
	1.60509417453279	-24.6670278319015 \\
	1.92611300943935	-23.7012054634436 \\
	2.24713184434591	-24.469084969449 \\
	2.56815067925247	-24.8867348342728 \\
	2.88916951415903	-26.7313966615265 \\
	3.21018834906559	-23.8765295899272 \\
	3.53120718397215	-24.422769695065 \\
	3.8522260188787	-23.7519767599219 \\
	4.17324485378526	-23.7838477560507 \\
	4.49426368869182	-20.1070540293602 \\
	4.81528252359838	-16.4320333462675 \\
	5.13630135850494	-5.85952229497303 \\
	5.4573201934115	-0.334326509003233 \\
	5.77833902831806	-1.11497900174113 \\
	6.09935786322461	-9.40450207340437 \\
	6.42037669813117	-17.5363759994132 \\
	6.74139553303773	-21.461614585098 \\
	7.06241436794429	-23.6217143431669 \\
	7.38343320285085	-25.6462028196862 \\
	7.70445203775741	-27.4451568957044 \\
	8.02547087266397	-27.6783302298637 \\
	8.34648970757053	-24.8163160573761 \\
	8.66750854247709	-23.7665424195428 \\
	8.98852737738364	-24.7742556425636 \\
	9.3095462122902	-21.146165101638 \\
	9.63056504719676	-23.0789335355151 \\
	9.95158388210332	-32.9033014142213 \\
	10.2726027170099	-26.6664018801621 \\
	10.5936215519164	-23.432503029009 \\
	10.914640386823	-28.7324683810203 \\
	11.2356592217296	-31.4191273643449 \\
	11.5566780566361	-24.8593897177413 \\
	11.8776968915427	-22.8113245019536 \\
	12.1987157264492	-23.7423289935668 \\
	12.5197345613558	-24.9216356587113 \\
	12.8407533962623	-25.7756407313235 \\
	13.1617722311689	-27.5810442042312 \\
	13.4827910660755	-26.8249859977423 \\
	13.803809900982	-26.6757475553122 \\
	14.1248287358886	-30.0611404205371 \\
	14.4458475707951	-30.2999275031628 \\
	14.7668664057017	-27.3571283657871 \\
	15.0878852406083	-28.3973631850874 \\
	15.4089040755148	-23.527379505061 \\
	15.7299229104214	-23.4567722160972 \\
	16.0509417453279	-24.9532389287162 \\
	16.3719605802345	-28.5908765816975 \\
	16.6929794151411	-29.8833111872802 \\
	17.0139982500476	-24.2578952393011 \\
	17.3350170849542	-24.2369242872695 \\
	17.6560359198607	-30.0571842412762 \\
	17.9770547547673	-26.8358487856504 \\
	18.2980735896738	-28.7555113839534 \\
	18.6190924245804	-28.5361956801343 \\
	18.940111259487	-23.8189949089156 \\
	19.2611300943935	-21.0288256462947 \\
	19.5821489293001	-22.9630825051246 \\
	19.9031677642066	-26.6342604138845 \\
	20.2241865991132	-24.3236449231652 \\
};
\addplot [semithick, blue]
table [row sep=\\]{%
	-20.5452054340198	-10.8986840359882 \\
	-20.2241865991132	-10.3750856237632 \\
	-19.9031677642066	-14.5466109138686 \\
	-19.5821489293001	-13.4209619080079 \\
	-19.2611300943935	-14.3898461468766 \\
	-18.940111259487	-13.278855826012 \\
	-18.6190924245804	-11.0113502449821 \\
	-18.2980735896738	-13.1132260421701 \\
	-17.9770547547673	-16.9446274454883 \\
	-17.6560359198607	-21.8337432409803 \\
	-17.3350170849542	-12.2725214363252 \\
	-17.0139982500476	-8.459548153749 \\
	-16.6929794151411	-9.12226469388631 \\
	-16.3719605802345	-11.4647120474058 \\
	-16.0509417453279	-8.42156054749831 \\
	-15.7299229104214	-9.33074800836652 \\
	-15.4089040755148	-9.29749369507038 \\
	-15.0878852406083	-9.55102964333831 \\
	-14.7668664057017	-8.85811035193378 \\
	-14.4458475707951	-12.4115033345554 \\
	-14.1248287358886	-12.1775599890979 \\
	-13.803809900982	-12.6281254099476 \\
	-13.4827910660755	-9.40598159836039 \\
	-13.1617722311689	-16.048738673075 \\
	-12.8407533962623	-13.3213544835168 \\
	-12.5197345613558	-11.2348404576616 \\
	-12.1987157264492	-9.14147404347011 \\
	-11.8776968915427	-12.6488603933113 \\
	-11.5566780566361	-10.9228880852857 \\
	-11.2356592217296	-8.64237083530544 \\
	-10.914640386823	-10.8929392511114 \\
	-10.5936215519164	-11.5463238597608 \\
	-10.2726027170099	-9.69067172596446 \\
	-9.95158388210332	-9.57277945779782 \\
	-9.63056504719676	-9.19386512860477 \\
	-9.3095462122902	-15.3927195701374 \\
	-8.98852737738364	-8.5974353928793 \\
	-8.66750854247709	-9.83784571957707 \\
	-8.34648970757053	-16.8164872291922 \\
	-8.02547087266397	-12.8378685508022 \\
	-7.70445203775741	-9.5902056267099 \\
	-7.38343320285085	-9.83635381629546 \\
	-7.06241436794429	-12.2637385489552 \\
	-6.74139553303773	-12.3012712030204 \\
	-6.42037669813117	-8.9251244056917 \\
	-6.09935786322461	-8.21486557952195 \\
	-5.77833902831806	-14.051323342564 \\
	-5.4573201934115	-8.55370337356821 \\
	-5.13630135850494	-10.3089692015937 \\
	-4.81528252359838	-10.087599476521 \\
	-4.49426368869182	-9.80012057358696 \\
	-4.17324485378526	-8.4041815989408 \\
	-3.8522260188787	-8.62668934055787 \\
	-3.53120718397215	-14.5088339000978 \\
	-3.21018834906559	-10.8372617030242 \\
	-2.88916951415903	-13.1335860470941 \\
	-2.56815067925247	-10.8090325652545 \\
	-2.24713184434591	-9.87568215365668 \\
	-1.92611300943935	-10.2814527806088 \\
	-1.60509417453279	-13.7339192175464 \\
	-1.28407533962623	-15.1699087547693 \\
	-0.963056504719676	-10.4164311492372 \\
	-0.642037669813117	-8.78244823421493 \\
	-0.321018834906559	-8.60482719723407 \\
	0	-9.82488998861953 \\
	0.321018834906559	-11.2191853475987 \\
	0.642037669813117	-15.489228576807 \\
	0.963056504719676	-13.9929248852226 \\
	1.28407533962623	-13.6261390362106 \\
	1.60509417453279	-9.83782186749882 \\
	1.92611300943935	-11.1077006516635 \\
	2.24713184434591	-9.84200415542919 \\
	2.56815067925247	-8.35929600443252 \\
	2.88916951415903	-11.0247782433553 \\
	3.21018834906559	-14.6345244515385 \\
	3.53120718397215	-11.2728682742046 \\
	3.8522260188787	-10.7104212572124 \\
	4.17324485378526	-11.5600941278637 \\
	4.49426368869182	-10.9684288678616 \\
	4.81528252359838	-10.3052095278871 \\
	5.13630135850494	-8.66183088758918 \\
	5.4573201934115	-1.58533128298047 \\
	5.77833902831806	-1.61876328740401 \\
	6.09935786322461	-6.47562928516615 \\
	6.42037669813117	-10.7296955932441 \\
	6.74139553303773	-9.88357133412205 \\
	7.06241436794429	-20.8293028001158 \\
	7.38343320285085	-9.62691791442607 \\
	7.70445203775741	-13.5799389412477 \\
	8.02547087266397	-13.4691677104865 \\
	8.34648970757053	-17.8152677816923 \\
	8.66750854247709	-11.6532946594733 \\
	8.98852737738364	-10.9634129012744 \\
	9.3095462122902	-12.0306035615257 \\
	9.63056504719676	-14.5306160629749 \\
	9.95158388210332	-13.2259923062251 \\
	10.2726027170099	-12.2798484650842 \\
	10.5936215519164	-10.1892384316939 \\
	10.914640386823	-8.79138930358741 \\
	11.2356592217296	-9.24769916308593 \\
	11.5566780566361	-13.7978205325932 \\
	11.8776968915427	-17.7955280349866 \\
	12.1987157264492	-12.2694160934029 \\
	12.5197345613558	-8.04659869156425 \\
	12.8407533962623	-9.56284409841922 \\
	13.1617722311689	-13.0597880823337 \\
	13.4827910660755	-9.49883485668078 \\
	13.803809900982	-8.60416471742561 \\
	14.1248287358886	-10.2828427748329 \\
	14.4458475707951	-14.1955239160063 \\
	14.7668664057017	-9.01715776514764 \\
	15.0878852406083	-12.6430959749883 \\
	15.4089040755148	-11.1512456552308 \\
	15.7299229104214	-10.4580439678555 \\
	16.0509417453279	-12.1459109026819 \\
	16.3719605802345	-18.4193634113305 \\
	16.6929794151411	-14.821146256442 \\
	17.0139982500476	-13.1668531088882 \\
	17.3350170849542	-9.53234168403918 \\
	17.6560359198607	-9.85872261303456 \\
	17.9770547547673	-11.0834741715497 \\
	18.2980735896738	-8.92243585483593 \\
	18.6190924245804	-9.29698411944283 \\
	18.940111259487	-10.6369683383711 \\
	19.2611300943935	-9.53142452151308 \\
	19.5821489293001	-10.4704935763214 \\
	19.9031677642066	-8.91278696721081 \\
	20.2241865991132	-9.60601501122781 \\
};
\addplot [semithick, color0]
table [row sep=\\]{%
	-20.5452054340198	-19.6624380572145 \\
	-20.2241865991132	-30.210710130762 \\
	-19.9031677642066	-23.5425429104158 \\
	-19.5821489293001	-27.1879735378788 \\
	-19.2611300943935	-24.7159903770007 \\
	-18.940111259487	-31.9879895408454 \\
	-18.6190924245804	-32.0748104062475 \\
	-18.2980735896738	-21.894590882382 \\
	-17.9770547547673	-20.1239413965455 \\
	-17.6560359198607	-23.1829779636576 \\
	-17.3350170849542	-20.9473034523374 \\
	-17.0139982500476	-21.084573125021 \\
	-16.6929794151411	-16.1250410793019 \\
	-16.3719605802345	-16.3237362846711 \\
	-16.0509417453279	-25.4072940565928 \\
	-15.7299229104214	-25.0739108607607 \\
	-15.4089040755148	-21.985562227228 \\
	-15.0878852406083	-23.3166789183932 \\
	-14.7668664057017	-22.2740527350705 \\
	-14.4458475707951	-20.0323162696141 \\
	-14.1248287358886	-20.74691296539 \\
	-13.803809900982	-20.8841037536092 \\
	-13.4827910660755	-18.4777543778081 \\
	-13.1617722311689	-22.1453483670065 \\
	-12.8407533962623	-22.6831871599692 \\
	-12.5197345613558	-23.2178403247845 \\
	-12.1987157264492	-23.4688025188004 \\
	-11.8776968915427	-19.5475095643631 \\
	-11.5566780566361	-16.5290415884467 \\
	-11.2356592217296	-18.8154176182507 \\
	-10.914640386823	-29.5968751513355 \\
	-10.5936215519164	-23.4596436076879 \\
	-10.2726027170099	-22.7280528317891 \\
	-9.95158388210332	-25.3359752904352 \\
	-9.63056504719676	-19.3026112508734 \\
	-9.3095462122902	-14.8003512976658 \\
	-8.98852737738364	-17.1721701166375 \\
	-8.66750854247709	-15.5052123255016 \\
	-8.34648970757053	-17.8908006517813 \\
	-8.02547087266397	-27.1264748673289 \\
	-7.70445203775741	-28.857100591454 \\
	-7.38343320285085	-27.179635379545 \\
	-7.06241436794429	-23.6498890930489 \\
	-6.74139553303773	-21.5301663304771 \\
	-6.42037669813117	-23.2286062540925 \\
	-6.09935786322461	-16.6393293458096 \\
	-5.77833902831806	-14.6391854027367 \\
	-5.4573201934115	-17.5919410273098 \\
	-5.13630135850494	-22.3700168135728 \\
	-4.81528252359838	-24.4095286805174 \\
	-4.49426368869182	-25.9197736012681 \\
	-4.17324485378526	-26.5671866502062 \\
	-3.8522260188787	-24.9457140513807 \\
	-3.53120718397215	-21.9295625916039 \\
	-3.21018834906559	-16.5623464239082 \\
	-2.88916951415903	-15.514009209661 \\
	-2.56815067925247	-20.8374450643997 \\
	-2.24713184434591	-20.3790104273174 \\
	-1.92611300943935	-14.7352476692161 \\
	-1.60509417453279	-15.519354681021 \\
	-1.28407533962623	-20.7802381407706 \\
	-0.963056504719676	-22.5958409258814 \\
	-0.642037669813117	-23.9568248977773 \\
	-0.321018834906559	-15.4699619455305 \\
	0	-14.409555252587 \\
	0.321018834906559	-18.2087636468713 \\
	0.642037669813117	-23.0666860201805 \\
	0.963056504719676	-24.3410628729214 \\
	1.28407533962623	-24.9518724872648 \\
	1.60509417453279	-22.2024783732306 \\
	1.92611300943935	-22.4071359654367 \\
	2.24713184434591	-24.9880676995856 \\
	2.56815067925247	-15.9308346106372 \\
	2.88916951415903	-15.3396663422805 \\
	3.21018834906559	-19.3019901696445 \\
	3.53120718397215	-22.7004214988324 \\
	3.8522260188787	-25.0798849060661 \\
	4.17324485378526	-23.9709709279473 \\
	4.49426368869182	-20.0583808160367 \\
	4.81528252359838	-16.1400791404918 \\
	5.13630135850494	-5.87707061006869 \\
	5.4573201934115	-0.337580148932136 \\
	5.77833902831806	-1.10626425477758 \\
	6.09935786322461	-9.28008703549895 \\
	6.42037669813117	-17.6986639659282 \\
	6.74139553303773	-21.3122168369559 \\
	7.06241436794429	-23.3467366193472 \\
	7.38343320285085	-26.7978292409337 \\
	7.70445203775741	-26.606782125531 \\
	8.02547087266397	-19.7230185869425 \\
	8.34648970757053	-14.9343984603406 \\
	8.66750854247709	-15.8074880792561 \\
	8.98852737738364	-20.5867691311371 \\
	9.3095462122902	-20.4239916212251 \\
	9.63056504719676	-24.1560670420909 \\
	9.95158388210332	-21.6923361485385 \\
	10.2726027170099	-26.4707961531011 \\
	10.5936215519164	-21.9222187664647 \\
	10.914640386823	-18.6540583533805 \\
	11.2356592217296	-15.8190576615436 \\
	11.5566780566361	-18.1021253084994 \\
	11.8776968915427	-22.8489699472519 \\
	12.1987157264492	-22.4824711267238 \\
	12.5197345613558	-18.8179637167694 \\
	12.8407533962623	-15.1687665158752 \\
	13.1617722311689	-16.5575279672883 \\
	13.4827910660755	-22.7822931237621 \\
	13.803809900982	-18.82590662852 \\
	14.1248287358886	-16.4830587330417 \\
	14.4458475707951	-20.5423823105306 \\
	14.7668664057017	-26.9257423134829 \\
	15.0878852406083	-27.3709924471312 \\
	15.4089040755148	-24.3176592674555 \\
	15.7299229104214	-23.4340397333798 \\
	16.0509417453279	-24.403610009585 \\
	16.3719605802345	-26.9225591689804 \\
	16.6929794151411	-19.6080359699152 \\
	17.0139982500476	-16.2389703126323 \\
	17.3350170849542	-18.0445047243567 \\
	17.6560359198607	-28.45161251418 \\
	17.9770547547673	-27.3912317701902 \\
	18.2980735896738	-33.4797672631896 \\
	18.6190924245804	-26.5865849789545 \\
	18.940111259487	-23.4063929896988 \\
	19.2611300943935	-20.401732968058 \\
	19.5821489293001	-18.7444385650638 \\
	19.9031677642066	-21.5002009276282 \\
	20.2241865991132	-18.1249106280926 \\
};
\addplot [semithick,color0]
table[row sep=crcr]{%
	-20.5452054340198	-19.6624299971701\\
	-20.2241865991132	-30.2106369648239\\
	-19.9031677642066	-23.5425718197353\\
	-19.5821489293001	-27.1879115218702\\
	-19.2611300943935	-24.7159837338356\\
	-18.940111259487	-31.988071202754\\
	-18.6190924245804	-32.0747120916299\\
	-18.2980735896738	-21.8945703166601\\
	-17.9770547547673	-20.1239494818414\\
	-17.6560359198607	-23.1829851239208\\
	-17.3350170849542	-20.9472970905612\\
	-17.0139982500476	-21.084577297835\\
	-16.6929794151411	-16.1250422386342\\
	-16.3719605802345	-16.3237381345266\\
	-16.0509417453279	-25.4072599861164\\
	-15.7299229104214	-25.0739358933378\\
	-15.4089040755148	-21.9855543239751\\
	-15.0878852406083	-23.316699752667\\
	-14.7668664057017	-22.2740359662634\\
	-14.4458475707951	-20.0323170459336\\
	-14.1248287358886	-20.7469241587986\\
	-13.803809900982	-20.8840882680654\\
	-13.4827910660755	-18.4777560835091\\
	-13.1617722311689	-22.1453641657656\\
	-12.8407533962623	-22.6832104735272\\
	-12.5197345613558	-23.2178169808724\\
	-12.1987157264492	-23.4688250287722\\
	-11.8776968915427	-19.5474975773968\\
	-11.5566780566361	-16.5290459912473\\
	-11.2356592217296	-18.8154158836719\\
	-10.914640386823	-29.5969361278075\\
	-10.5936215519164	-23.4596271784504\\
	-10.2726027170099	-22.7280427998232\\
	-9.95158388210332	-25.3359848837729\\
	-9.63056504719676	-19.3026125436757\\
	-9.3095462122902	-14.8003486562168\\
	-8.98852737738364	-17.1721760282668\\
	-8.66750854247709	-15.505216658295\\
	-8.34648970757053	-17.890795664027\\
	-8.02547087266397	-27.1264295195925\\
	-7.70445203775741	-28.8570510138581\\
	-7.38343320285085	-27.1796461591599\\
	-7.06241436794429	-23.6499189060591\\
	-6.74139553303773	-21.5301553943878\\
	-6.42037669813117	-23.2285871142625\\
	-6.09935786322461	-16.6393307325065\\
	-5.77833902831806	-14.6391868770331\\
	-5.4573201934115	-17.591933479499\\
	-5.13630135850494	-22.3700275636878\\
	-4.81528252359838	-24.4095521712629\\
	-4.49426368869182	-25.9198224117955\\
	-4.17324485378526	-26.5671311113484\\
	-3.8522260188787	-24.94575489837\\
	-3.53120718397215	-21.9295446358742\\
	-3.21018834906559	-16.5623509860459\\
	-2.88916951415903	-15.5140072869312\\
	-2.56815067925247	-20.8374384029101\\
	-2.24713184434591	-20.3790216529513\\
	-1.92611300943935	-14.7352434342076\\
	-1.60509417453279	-15.5193580234184\\
	-1.28407533962623	-20.7802228891244\\
	-0.963056504719676	-22.5958604402746\\
	-0.642037669813117	-23.9568239701132\\
	-0.321018834906559	-15.4699659824052\\
	0	-14.4095518320338\\
	0.321018834906559	-18.2087628301606\\
	0.642037669813117	-23.0667124030645\\
	0.963056504719676	-24.3410375829314\\
	1.28407533962623	-24.9519122602768\\
	1.60509417453279	-22.2024640563115\\
	1.92611300943935	-22.4071398660313\\
	2.24713184434591	-24.9880970805891\\
	2.56815067925247	-15.9308360928612\\
	2.88916951415903	-15.3396620208025\\
	3.21018834906559	-19.3020009790119\\
	3.53120718397215	-22.7003991539876\\
	3.8522260188787	-25.0799092958232\\
	4.17324485378526	-23.970993775192\\
	4.49426368869182	-20.058372688111\\
	4.81528252359838	-16.1400741730983\\
	5.13630135850494	-5.87706980475469\\
	5.4573201934115	-0.33757988480308\\
	5.77833902831806	-1.10626399132087\\
	6.09935786322461	-9.28008567420324\\
	6.42037669813117	-17.69865640073\\
	6.74139553303773	-21.3122034609332\\
	7.06241436794429	-23.3467453636938\\
	7.38343320285085	-26.7978133102314\\
	7.70445203775741	-26.6068292899333\\
	8.02547087266397	-19.7230304977189\\
	8.34648970757053	-14.9343941794113\\
	8.66750854247709	-15.8074908632106\\
	8.98852737738364	-20.5867728104169\\
	9.3095462122902	-20.4239790822664\\
	9.63056504719676	-24.1560968478219\\
	9.95158388210332	-21.6923164359734\\
	10.2726027170099	-26.4707374194798\\
	10.5936215519164	-21.9222217058635\\
	10.914640386823	-18.6540611714615\\
	11.2356592217296	-15.8190528760536\\
	11.5566780566361	-18.102132765264\\
	11.8776968915427	-22.8489860659131\\
	12.1987157264492	-22.4824475689625\\
	12.5197345613558	-18.8179659384519\\
	12.8407533962623	-15.1687685915658\\
	13.1617722311689	-16.5575221684107\\
	13.4827910660755	-22.7823134151069\\
	13.803809900982	-18.82590045218\\
	14.1248287358886	-16.4830567835475\\
	14.4458475707951	-20.5423933859579\\
	14.7668664057017	-26.9257575868576\\
	15.0878852406083	-27.3710640087358\\
	15.4089040755148	-24.3176365300249\\
	15.7299229104214	-23.4340390670101\\
	16.0509417453279	-24.4036450519735\\
	16.3719605802345	-26.9225562113596\\
	16.6929794151411	-19.6080238134163\\
	17.0139982500476	-16.2389739745422\\
	17.3350170849542	-18.0445028097969\\
	17.6560359198607	-28.4516746645848\\
	17.9770547547673	-27.3911608734109\\
	18.2980735896738	-33.4800060271554\\
	18.6190924245804	-26.5865707946022\\
	18.940111259487	-23.4064206914501\\
	19.2611300943935	-20.4017343847772\\
	19.5821489293001	-18.744429623758\\
	19.9031677642066	-21.5002132162589\\
	20.2241865991132	-18.1249188129007\\
};

\addplot [semithick,orange]
table[row sep=crcr]{%
	-20.5452054340198	-13.8573115509201\\
	-20.2241865991132	-21.460453688732\\
	-19.9031677642066	-14.1422392027355\\
	-19.5821489293001	-13.7811842312284\\
	-19.2611300943935	-15.5137531414616\\
	-18.940111259487	-16.9955081336195\\
	-18.6190924245804	-13.9033976147048\\
	-18.2980735896738	-14.6746915150187\\
	-17.9770547547673	-18.5951990398667\\
	-17.6560359198607	-15.1406760244668\\
	-17.3350170849542	-12.3853745456528\\
	-17.0139982500476	-11.0435205571064\\
	-16.6929794151411	-9.91553850461185\\
	-16.3719605802345	-12.7437885057477\\
	-16.0509417453279	-11.1606959256531\\
	-15.7299229104214	-16.4496249717922\\
	-15.4089040755148	-14.836535926827\\
	-15.0878852406083	-12.2146626345244\\
	-14.7668664057017	-9.64078529051773\\
	-14.4458475707951	-9.3209698050893\\
	-14.1248287358886	-13.0859281844523\\
	-13.803809900982	-10.7931628034284\\
	-13.4827910660755	-8.07586430680933\\
	-13.1617722311689	-11.3385954107999\\
	-12.8407533962623	-16.1223519012309\\
	-12.5197345613558	-16.0757612889758\\
	-12.1987157264492	-12.556397940119\\
	-11.8776968915427	-13.0670780680553\\
	-11.5566780566361	-10.2580519571987\\
	-11.2356592217296	-10.7360492479288\\
	-10.914640386823	-14.3446790967068\\
	-10.5936215519164	-14.3364211567514\\
	-10.2726027170099	-12.0403454687472\\
	-9.95158388210332	-13.4577926619652\\
	-9.63056504719676	-11.6265687626456\\
	-9.3095462122902	-10.3729960683753\\
	-8.98852737738364	-12.1091881945136\\
	-8.66750854247709	-17.1776736082451\\
	-8.34648970757053	-17.7512300474676\\
	-8.02547087266397	-20.1806666454532\\
	-7.70445203775741	-11.0057659397027\\
	-7.38343320285085	-9.48962239636156\\
	-7.06241436794429	-10.7818239052923\\
	-6.74139553303773	-14.5439327855609\\
	-6.42037669813117	-9.67561296151632\\
	-6.09935786322461	-9.13678896531068\\
	-5.77833902831806	-14.9780137458989\\
	-5.4573201934115	-12.1576739654856\\
	-5.13630135850494	-14.235746079799\\
	-4.81528252359838	-10.9084194341861\\
	-4.49426368869182	-10.0278096484437\\
	-4.17324485378526	-9.97814226802427\\
	-3.8522260188787	-11.0224880222753\\
	-3.53120718397215	-25.0040195253305\\
	-3.21018834906559	-14.5564561826663\\
	-2.88916951415903	-14.9009791493582\\
	-2.56815067925247	-14.0214720063292\\
	-2.24713184434591	-11.9382032350005\\
	-1.92611300943935	-8.26504983189771\\
	-1.60509417453279	-9.3633011002382\\
	-1.28407533962623	-15.1274667924946\\
	-0.963056504719676	-13.9665689592683\\
	-0.642037669813117	-12.4219068719608\\
	-0.321018834906559	-10.6373651342402\\
	0	-10.1954342434019\\
	0.321018834906559	-10.3395008969726\\
	0.642037669813117	-13.516024868589\\
	0.963056504719676	-13.3031691747425\\
	1.28407533962623	-15.6236289021073\\
	1.60509417453279	-11.2775518056878\\
	1.92611300943935	-13.5170068655908\\
	2.24713184434591	-12.0895714051946\\
	2.56815067925247	-9.42349917080559\\
	2.88916951415903	-9.82594929588253\\
	3.21018834906559	-12.05632971021\\
	3.53120718397215	-15.1842685812021\\
	3.8522260188787	-13.1277583226597\\
	4.17324485378526	-11.3227381574197\\
	4.49426368869182	-10.7832087816036\\
	4.81528252359838	-10.5505216755594\\
	5.13630135850494	-7.0850436614737\\
	5.4573201934115	-0.665028984366416\\
	5.77833902831806	-1.15447074618535\\
	6.09935786322461	-7.42559130748942\\
	6.42037669813117	-12.8608413421614\\
	6.74139553303773	-13.9122777529099\\
	7.06241436794429	-11.9569393100306\\
	7.38343320285085	-9.26734737935819\\
	7.70445203775741	-11.8003396438775\\
	8.02547087266397	-20.5444006928714\\
	8.34648970757053	-12.2015589890197\\
	8.66750854247709	-12.6255711844678\\
	8.98852737738364	-17.3194010704392\\
	9.3095462122902	-16.5524462233006\\
	9.63056504719676	-12.7917049333657\\
	9.95158388210332	-11.8732209274041\\
	10.2726027170099	-13.4357915464175\\
	10.5936215519164	-12.378909037612\\
	10.914640386823	-12.5296170999242\\
	11.2356592217296	-8.67492429165912\\
	11.5566780566361	-9.68826789109499\\
	11.8776968915427	-11.8829427561327\\
	12.1987157264492	-11.3499141504747\\
	12.5197345613558	-8.43427145753719\\
	12.8407533962623	-8.20756125723456\\
	13.1617722311689	-13.1394403360739\\
	13.4827910660755	-11.9776788909065\\
	13.803809900982	-13.7943868245654\\
	14.1248287358886	-10.9358261554847\\
	14.4458475707951	-11.4843030957742\\
	14.7668664057017	-11.3261557641113\\
	15.0878852406083	-20.1899882060942\\
	15.4089040755148	-13.5977409253566\\
	15.7299229104214	-17.38086855298\\
	16.0509417453279	-19.1213704752203\\
	16.3719605802345	-14.6602671135643\\
	16.6929794151411	-19.1595947281813\\
	17.0139982500476	-11.2533105324857\\
	17.3350170849542	-11.8553738887379\\
	17.6560359198607	-11.6964066433591\\
	17.9770547547673	-12.1295267789717\\
	18.2980735896738	-11.1762399576933\\
	18.6190924245804	-11.0634660461808\\
	18.940111259487	-10.6864354335713\\
	19.2611300943935	-9.83745645485722\\
	19.5821489293001	-11.7369609586455\\
	19.9031677642066	-10.393686929983\\
	20.2241865991132	-8.99620782842172\\
};

\addplot [semithick,color2]
table[row sep=crcr]{%
	-20.5452054340198	-23.1869045722359\\
	-20.2241865991132	-21.5519079596709\\
	-19.9031677642066	-23.9764786539805\\
	-19.5821489293001	-21.2443831180805\\
	-19.2611300943935	-28.0431274323734\\
	-18.940111259487	-20.2856798984509\\
	-18.6190924245804	-19.9522301287788\\
	-18.2980735896738	-18.0864799826252\\
	-17.9770547547673	-19.6584541267234\\
	-17.6560359198607	-24.4569120843307\\
	-17.3350170849542	-21.9114608975946\\
	-17.0139982500476	-22.2928333008685\\
	-16.6929794151411	-18.5188055469021\\
	-16.3719605802345	-20.8419521673062\\
	-16.0509417453279	-25.1598715446685\\
	-15.7299229104214	-25.8944029998745\\
	-15.4089040755148	-21.0806612283417\\
	-15.0878852406083	-25.777065135857\\
	-14.7668664057017	-23.5479958264277\\
	-14.4458475707951	-20.3734563104491\\
	-14.1248287358886	-22.1854040203463\\
	-13.803809900982	-25.7401463030838\\
	-13.4827910660755	-26.0056639334382\\
	-13.1617722311689	-26.4775014518286\\
	-12.8407533962623	-21.8837115162706\\
	-12.5197345613558	-21.3069011452504\\
	-12.1987157264492	-31.9367526347852\\
	-11.8776968915427	-21.5743453987155\\
	-11.5566780566361	-17.7184850839076\\
	-11.2356592217296	-22.6725767644921\\
	-10.914640386823	-21.5982056408642\\
	-10.5936215519164	-22.3660621820057\\
	-10.2726027170099	-22.3159252738835\\
	-9.95158388210332	-24.6964670249987\\
	-9.63056504719676	-24.1273202106944\\
	-9.3095462122902	-18.5670140581847\\
	-8.98852737738364	-23.1756148409096\\
	-8.66750854247709	-17.8518008147801\\
	-8.34648970757053	-23.3730585520004\\
	-8.02547087266397	-22.1163518416545\\
	-7.70445203775741	-23.948649447067\\
	-7.38343320285085	-21.6852769678467\\
	-7.06241436794429	-24.4886720923169\\
	-6.74139553303773	-28.4853982899802\\
	-6.42037669813117	-19.4773981157976\\
	-6.09935786322461	-21.4182153236711\\
	-5.77833902831806	-18.9524897630274\\
	-5.4573201934115	-22.7032916970606\\
	-5.13630135850494	-31.2196741669381\\
	-4.81528252359838	-20.9344761632047\\
	-4.49426368869182	-25.8952388579988\\
	-4.17324485378526	-20.1009186988503\\
	-3.8522260188787	-22.1550565444144\\
	-3.53120718397215	-25.4495385221846\\
	-3.21018834906559	-22.5117449518073\\
	-2.88916951415903	-21.5419502812403\\
	-2.56815067925247	-24.9117515181803\\
	-2.24713184434591	-28.4073909614499\\
	-1.92611300943935	-19.6690212828818\\
	-1.60509417453279	-21.6913898191777\\
	-1.28407533962623	-22.7786019430245\\
	-0.963056504719676	-24.6163247776298\\
	-0.642037669813117	-21.8859376441104\\
	-0.321018834906559	-22.0195262085704\\
	0	-18.7309512350447\\
	0.321018834906559	-21.5915149210163\\
	0.642037669813117	-24.5102937626734\\
	0.963056504719676	-20.464856790136\\
	1.28407533962623	-20.4910462826889\\
	1.60509417453279	-25.9421083057397\\
	1.92611300943935	-21.8056036239913\\
	2.24713184434591	-20.796968466165\\
	2.56815067925247	-32.3990194294463\\
	2.88916951415903	-26.3720100488334\\
	3.21018834906559	-21.1429882258971\\
	3.53120718397215	-22.2299302648235\\
	3.8522260188787	-19.8486457724893\\
	4.17324485378526	-21.3851193187491\\
	4.49426368869182	-22.8134562022611\\
	4.81528252359838	-16.4471589388603\\
	5.13630135850494	-5.80223713248842\\
	5.4573201934115	-0.33091467240698\\
	5.77833902831806	-1.10376953560237\\
	6.09935786322461	-9.22823700784093\\
	6.42037669813117	-17.0452713658885\\
	6.74139553303773	-33.5711190455922\\
	7.06241436794429	-22.023236960641\\
	7.38343320285085	-22.8499255243984\\
	7.70445203775741	-22.2461957195104\\
	8.02547087266397	-21.5337304950878\\
	8.34648970757053	-23.8838209615291\\
	8.66750854247709	-21.7618414672291\\
	8.98852737738364	-26.5465566067663\\
	9.3095462122902	-20.4824122560176\\
	9.63056504719676	-20.9377527878825\\
	9.95158388210332	-31.7730835142896\\
	10.2726027170099	-24.9310165435711\\
	10.5936215519164	-27.7630662476508\\
	10.914640386823	-24.8486911993268\\
	11.2356592217296	-27.9400630195339\\
	11.5566780566361	-25.7323134307093\\
	11.8776968915427	-21.0132181513888\\
	12.1987157264492	-22.6158837601321\\
	12.5197345613558	-26.5405571324608\\
	12.8407533962623	-23.2601003233339\\
	13.1617722311689	-26.5346922343473\\
	13.4827910660755	-24.3488025640078\\
	13.803809900982	-26.0517513681869\\
	14.1248287358886	-28.9249826052753\\
	14.4458475707951	-24.3559764258657\\
	14.7668664057017	-23.2143170599842\\
	15.0878852406083	-22.2526733484748\\
	15.4089040755148	-20.4178131915778\\
	15.7299229104214	-31.3447095340586\\
	16.0509417453279	-29.233051395662\\
	16.3719605802345	-22.38438774253\\
	16.6929794151411	-23.6411302765818\\
	17.0139982500476	-25.5772795180744\\
	17.3350170849542	-28.7183475084975\\
	17.6560359198607	-28.1395050281164\\
	17.9770547547673	-21.1641038304394\\
	18.2980735896738	-24.3547950211417\\
	18.6190924245804	-22.8038371959336\\
	18.940111259487	-22.8397090942471\\
	19.2611300943935	-22.9763007554677\\
	19.5821489293001	-25.5426769408294\\
	19.9031677642066	-20.9758533477764\\
	20.2241865991132	-20.899147182643\\
};
\addplot [semithick, color1]
table [row sep=\\]{%
	-20.5452054340198	-61.5220289653563 \\
	-20.2241865991132	-61.5220289653563 \\
	-19.9031677642066	-61.5220289653563 \\
	-19.5821489293001	-61.5220289653563 \\
	-19.2611300943935	-61.5220289653563 \\
	-18.940111259487	-61.5220289653563 \\
	-18.6190924245804	-61.5220289653563 \\
	-18.2980735896738	-61.5220289653563 \\
	-17.9770547547673	-61.5220289653563 \\
	-17.6560359198607	-61.5220289653563 \\
	-17.3350170849542	-61.5220289653563 \\
	-17.0139982500476	-61.5220289653563 \\
	-16.6929794151411	-61.5220289653563 \\
	-16.3719605802345	-61.5220289653563 \\
	-16.0509417453279	-61.5220289653563 \\
	-15.7299229104214	-61.5220289653563 \\
	-15.4089040755148	-61.5220289653563 \\
	-15.0878852406083	-61.5220289653563 \\
	-14.7668664057017	-61.5220289653563 \\
	-14.4458475707951	-61.5220289653563 \\
	-14.1248287358886	-61.5220289653563 \\
	-13.803809900982	-61.5220289653563 \\
	-13.4827910660755	-61.5220289653563 \\
	-13.1617722311689	-61.5220289653563 \\
	-12.8407533962623	-61.5220289653563 \\
	-12.5197345613558	-61.5220289653563 \\
	-12.1987157264492	-61.5220289653563 \\
	-11.8776968915427	-61.5220289653563 \\
	-11.5566780566361	-61.5220289653563 \\
	-11.2356592217296	-61.5220289653563 \\
	-10.914640386823	-61.5220289653563 \\
	-10.5936215519164	-61.5220289653563 \\
	-10.2726027170099	-61.5220289653563 \\
	-9.95158388210332	-61.5220289653563 \\
	-9.63056504719676	-61.5220289653563 \\
	-9.3095462122902	-61.5220289653563 \\
	-8.98852737738364	-61.5220289653563 \\
	-8.66750854247709	-61.5220289653563 \\
	-8.34648970757053	-61.5220289653563 \\
	-8.02547087266397	-61.5220289653563 \\
	-7.70445203775741	-61.5220289653563 \\
	-7.38343320285085	-61.5220289653563 \\
	-7.06241436794429	-61.5220289653563 \\
	-6.74139553303773	-61.5220289653563 \\
	-6.42037669813117	-61.5220289653563 \\
	-6.09935786322461	-61.5220289653563 \\
	-5.77833902831806	-61.5220289653563 \\
	-5.4573201934115	-61.5220289653563 \\
	-5.13630135850494	-61.5220289653563 \\
	-4.81528252359838	-61.5220289653563 \\
	-4.49426368869182	-61.5220289653563 \\
	-4.17324485378526	-61.5220289653563 \\
	-3.8522260188787	-61.5220289653563 \\
	-3.53120718397215	-61.5220289653563 \\
	-3.21018834906559	-61.5220289653563 \\
	-2.88916951415903	-61.5220289653563 \\
	-2.56815067925247	-61.5220289653563 \\
	-2.24713184434591	-61.5220289653563 \\
	-1.92611300943935	-61.5220289653563 \\
	-1.60509417453279	-61.5220289653563 \\
	-1.28407533962623	-61.5220289653563 \\
	-0.963056504719676	-61.5220289653563 \\
	-0.642037669813117	-61.5220289653563 \\
	-0.321018834906559	-61.5220289653563 \\
	0	-61.5220289653563 \\
	0.321018834906559	-61.5220289653563 \\
	0.642037669813117	-61.5220289653563 \\
	0.963056504719676	-61.5220289653563 \\
	1.28407533962623	-61.5220289653563 \\
	1.60509417453279	-61.5220289653563 \\
	1.92611300943935	-61.5220289653563 \\
	2.24713184434591	-61.5220289653563 \\
	2.56815067925247	-61.5220289653563 \\
	2.88916951415903	-61.5220289653563 \\
	3.21018834906559	-61.5220289653563 \\
	3.53120718397215	-61.5220289653563 \\
	3.8522260188787	-61.5220289653563 \\
	4.17324485378526	-61.5220289653563 \\
	4.49426368869182	-61.5220289653563 \\
	4.81528252359838	-18.1151236980752 \\
	5.13630135850494	-10.8319535778668 \\
	5.4573201934115	-2.90977679205268 \\
	5.77833902831806	-2.59489722814345 \\
	6.09935786322461	-9.78616504352075 \\
	6.42037669813117	-21.5475265833175 \\
	6.74139553303773	-61.5220289653563 \\
	7.06241436794429	-61.5220289653563 \\
	7.38343320285085	-61.5220289653563 \\
	7.70445203775741	-61.5220289653563 \\
	8.02547087266397	-61.5220289653563 \\
	8.34648970757053	-61.5220289653563 \\
	8.66750854247709	-61.5220289653563 \\
	8.98852737738364	-61.5220289653563 \\
	9.3095462122902	-61.5220289653563 \\
	9.63056504719676	-61.5220289653563 \\
	9.95158388210332	-61.5220289653563 \\
	10.2726027170099	-61.5220289653563 \\
	10.5936215519164	-61.5220289653563 \\
	10.914640386823	-61.5220289653563 \\
	11.2356592217296	-61.5220289653563 \\
	11.5566780566361	-61.5220289653563 \\
	11.8776968915427	-61.5220289653563 \\
	12.1987157264492	-61.5220289653563 \\
	12.5197345613558	-61.5220289653563 \\
	12.8407533962623	-61.5220289653563 \\
	13.1617722311689	-61.5220289653563 \\
	13.4827910660755	-61.5220289653563 \\
	13.803809900982	-61.5220289653563 \\
	14.1248287358886	-61.5220289653563 \\
	14.4458475707951	-61.5220289653563 \\
	14.7668664057017	-61.5220289653563 \\
	15.0878852406083	-61.5220289653563 \\
	15.4089040755148	-61.5220289653563 \\
	15.7299229104214	-61.5220289653563 \\
	16.0509417453279	-61.5220289653563 \\
	16.3719605802345	-61.5220289653563 \\
	16.6929794151411	-61.5220289653563 \\
	17.0139982500476	-61.5220289653563 \\
	17.3350170849542	-61.5220289653563 \\
	17.6560359198607	-61.5220289653563 \\
	17.9770547547673	-61.5220289653563 \\
	18.2980735896738	-61.5220289653563 \\
	18.6190924245804	-61.5220289653563 \\
	18.940111259487	-61.5220289653563 \\
	19.2611300943935	-61.5220289653563 \\
	19.5821489293001	-61.5220289653563 \\
	19.9031677642066	-61.5220289653563 \\
	20.2241865991132	-61.5220289653563 \\
};

\end{axis}

\end{tikzpicture}

%% file: conclusion.tex
\section{Conclusion}

In this paper, novel NN-based methods for effective mutual interference mitigation and denoising in the context of automotive radar sensors have been presented. Noteworthy is the usage of (complex-valued) spectrograms from different steps in the range-Doppler signal processing as network inputs and their suitability to be processed by CNNs.

An extensive simulation framework was used for data generation, training and evaluation. The most promising model architecture was then compared to a small selection of well-known conventional interference mitigation techniques. It was shown that the CNN-based model is capable of preserving the object peaks, while suppressing noise and interference by several orders of magnitude compared to conventional methods. However, its performance may be less robust, especially when considering the distortion of object peak values. Furthermore, the achieved amount of suppression indicates an implicit peak detection capability, which is of course attributed to the use of clean training data.

The most important issue in the future is to analyze the generalization capability of the architectures to real-world data. In addition, we would like to investigate the potential of temporal information in range-Doppler processing using NNs.